\documentclass[aps,prb,twocolumn,superscriptaddress,showpacs,floatfix,longbibliography,10pt]{revtex4-2}
\usepackage[english]{babel}
\usepackage{fullpage}
\usepackage{amssymb}
\usepackage{graphicx,color}
\usepackage{bbm}
\usepackage{multirow}
\usepackage{bm}
\usepackage{mathrsfs}
\usepackage{commath}
\usepackage{stmaryrd}
\usepackage{color}
\usepackage{subfigure}
\usepackage{comment}
\usepackage[utf8]{inputenc}
\usepackage[normalem]{ulem}
\usepackage{amsmath}
\usepackage{graphicx}
\usepackage{float}
\usepackage{verbatim}
\usepackage{esint}
\usepackage{epstopdf}
\usepackage{braket}
\usepackage[ruled,lined]{algorithm2e}

\usepackage{hyperref}
\hypersetup{
  colorlinks   = true,    
  urlcolor     = blue,    
  linkcolor    = blue,    
  citecolor    = blue     
}

\makeatletter
\@ifundefined{textcolor}{}
{%
 \definecolor{BLACK}{gray}{0}
 \definecolor{WHITe}{gray}{1}
 \definecolor{ReD}{rgb}{1,0,0}
 \definecolor{GReeN}{rgb}{0,1,0}
 \definecolor{BLUe}{rgb}{0,0,1}
 \definecolor{CYAN}{cmyk}{1,0,0,0}
 \definecolor{MAGeNTA}{cmyk}{0,1,0,0}
  \definecolor{PURPLE}{cmyk}{0.4,0.8,0,0}
 \definecolor{YeLLOW}{cmyk}{0,0,1,0}
}

\def\d{{\rm d}}
\def\R{{\rm R}}
\def\q{{\rm q}}
\def\c{{\rm c}}

\def\hS{\hat{S}}
\def\hb{\hat{b}}
\def\hc{\hat{c}}
\def\hn{\hat{n}}
\def\htheta{\hat{\theta}}
\def\hphi{\hat{\phi}}
\def\hpsi{\hat{\psi}}

\def\hrho{\hat{\rho}}

\usepackage{epsfig}
\usepackage{dcolumn}
\usepackage{bm}

\makeatother

\begin{document}

\title{The generic Mott transition in the sine-Gordon model through an embedded worm algorithm}

\author{Oscar Bouverot-Dupuis}
\affiliation{Universit\'{e} Paris Saclay, CNRS, LPTMS, 91405, Orsay, France}
\affiliation{IPhT, CNRS, CEA, Universit\'{e} Paris Saclay, 91191 Gif-sur-Yvette, France}

\author{Laura Foini}
\affiliation{IPhT, CNRS, CEA, Universit\'{e} Paris Saclay, 91191 Gif-sur-Yvette, France}

\author{Alberto Rosso}
\affiliation{Universit\'{e} Paris Saclay, CNRS, LPTMS, 91405, Orsay, France}

\begin{abstract}
The generic Mott transition in one-dimensional quantum systems can be described by the $(1+1)D$ sine-Gordon model with a tilt via bosonization. Because the configuration space of the sine-Gordon model separates into distinct topological sectors, standard local Monte Carlo schemes are limited to very small system sizes ($\sim 15 \times 15$). To overcome this limitation, we introduce the smooth worm (SmoWo) Monte Carlo algorithm which enlarges the configuration space to allow smooth transitions between topological sectors. The method combines worm updates with event-chain Monte Carlo moves. We explicitly prove its validity and quantify its performance. Thanks to the substantial acceleration achieved by the SmoWo algorithm, we are able to simulate large system sizes up to $256 \times 256$, providing a precise picture of the different phases and critical behaviour of the sine-Gordon model.
\end{abstract}

\date{\today}

\maketitle

\tableofcontents

\section{Introduction}
Over the past decades, the Mott transition \cite{Mott_1968,Fisher_1989_SI,Giamarchi_1997_Mott1D,Mott_2004} describing metal-insulator transitions driven by strong electronic interactions has attracted a great deal of interest. While in dimensions greater than one, a comprehensive theoretical framework remains elusive despite significant numerical efforts \cite{Georges_1996,Imada_1998}, the one-dimensional (1D) case has seen major advances. Numerically, the Bose Hubbard model has become a paradigmatic model to observe the Mott transition with quantum Monte Carlo algorithms \cite{Batrouni_1990_BoseHubbard_MC} and the density matrix renormalization group (DMRG) \cite{Kuhner_2000_BoseHubbard_DMRG}. Analytically, apart from strong-coupling expansions of the Bose Hubbard model \cite{Elstner_1999_BoseHubbard_12order}, most results rely on the description of the transition through the sine–Gordon model via bosonization \cite{Haldane_1981,Giamarchi,von_Delft_bosonization,Cazalilla_2004_harmonic_fluid,NdupuisCMUG2}. It was thus established that 1D Mott transitions fall into two distinct universality classes: varying the interaction strength at fixed commensurate densities leads to a Berezinskii--Kosterlitz--Thouless (BKT) transition, while doping the system gives rise to the \emph{generic} (or Mott-$\delta$) transition. The latter is also well-known for describing commensurate-incommensurate transitions in uniaxial surface structures \cite{Pokrovsky_1979,Schulz_1980_CI,Haldane_1983} and related systems \cite{Pokrovsky_1983_long_range_CI,Radzihovsky_2006_vortex_field}. Even within the sine-Gordon picture, however, its analysis proved non-trivial as a complete picture of the transition only came through the combination of many integrability \cite{Yamamoto_1983_sG,Haldane_1983,Papa_2001_sG} and field-theoretical techniques \cite{Schulz_1980_CI,Giamarchi_1997_Mott1D,Horovitz_1983_CI_RG,Aristov_2002_sG}.

In this work, we propose a numerical study of the 1D generic Mott transition. Rather than simulating a microscopic quantum Hamiltonian as is done in quantum Monte Carlo studies, we focus on the bosonized picture provided by the sine-Gordon model. This enables direct comparison with analytical results on the same model and serves as a starting point for Monte Carlo simulations of more complicated bosonized systems. We work in the grand-canonical ensemble, thereby complementing the canonical-ensemble algorithm of Ref.~\cite{bouvdup_2025_bosonized1d}. In this setting, a Monte Carlo simulation of the sine-Gordon model must sample its different topological sectors. Since these sectors are widely separated in configuration space, standard Monte Carlo schemes are limited to very small system sizes. To overcome this limitation, we introduce the smooth worm algorithm (SmoWo) which operates in an enlarged configuration space where topological sectors are smoothly connected. The algorithm combines updates from the worm algorithm \cite{Prokofiev_1998_worm,Prokofiev_2001_worm} with local moves of the event-chain Monte Carlo (ECMC) scheme \cite{Michel2014GenECMC,krauth2021ECMC}. The worm algorithm is designed to sample loop models and has already proven effective to study Mott transitions in coarse-grained $(2+1)$D models \cite{Alet_2003_geometricWA,Alet_2003_directedWA,Alet_2004_CI,Prokofiev_2004_SI_dis}. In the SmoWo algorithm, the worm updates are further accelerated by the ECMC which performs persistent and non-reversible moves to smooth the worm throughout its construction.

The paper is organized as follows. Section~\ref{sec:model} reviews the mapping of 1D systems onto the sine-Gordon model via bosonization, and then describes various lattice representations of the model. Section~\ref{sec:MC_algorithm} introduces the SmoWo algorithm and evaluates its performance. Section~\ref{sec:observables} defines the observables of interest and outlines their computation within the SmoWo algorithm. Section~\ref{sec:numerical_results} presents large-scale numerical results, and Section~\ref{sec:conclusion} concludes. Additional details on bosonization, a proof of the algorithm's validity and implementation details can be found in Appendixes~\ref{app:Bosonization_subtleties}--\ref{app:ECMC_in_practice}.

\section{Model}
\label{sec:model}
We wish to study the generic Mott transition in 1D quantum systems. A system that exhibits such a transition is the following tight-binding model for spinless fermions hopping on $N$ sites with lattice spacing $a$,
\begin{align}\label{eq:H_fermion}
    \hat{H}_{\rm fermions}&=\sum_{j=1}^N -t\left(\hc^\dagger_{j+1} \hc_j + \hc^\dagger_j \hc_{j+1}\right)\nonumber\\
    +& V\left(\hn_j-\frac{1}{2}\right)\left(\hn_{j+1}-\frac{1}{2}\right)- \mu \hn_j,
\end{align}
where $\mu$ is the chemical potential, $t$ the exchange integral and $V$ the amplitude of the nearest-neighbour interaction. This Hamiltonian can also describe an XXZ spin chain in a constant magnetic field $h=\mu$ as the Jordan-Wigner transformation maps it onto
\begin{align}\label{eq:H_spin}
    \hat{H}_{\rm spin}=\sum_{j=1}^N & - 2 t\left( \hS_j^x \hS_{j+1}^x + \hS_j^y \hS_{j+1}^y\right)\nonumber\\
    & +  V \hS_j^z \hS_{j+1}^z - \mu \hS_j^z.
\end{align}
Intuitively, particles have been replaced by up spins and holes by down spins. 

\subsection{Bosonization}
\begin{figure}[h!]
    \centering
    \includegraphics[width=0.8\linewidth]{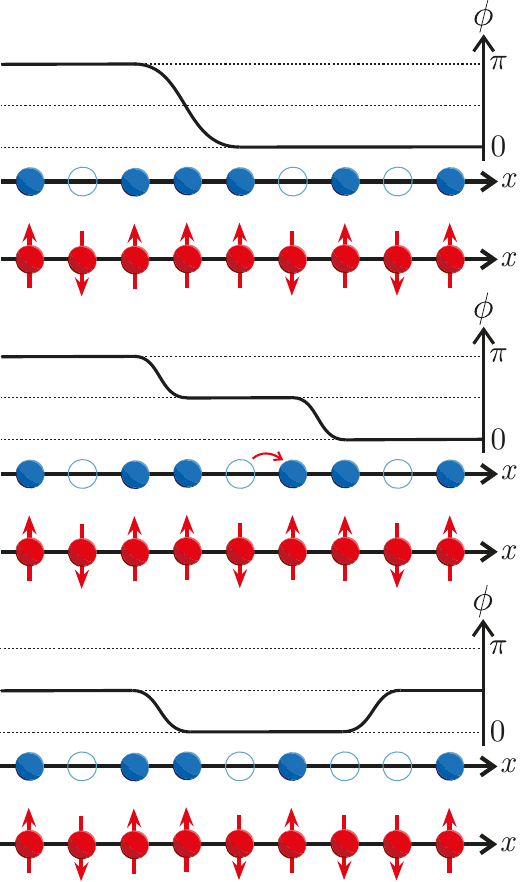}
    \caption{Top: In the bosonized picture, particles (or up spins) sitting above the density $\rho_0$ create three consecutive particles and are represented by kinks of height $-\pi$. An extra hole (or spin down) creates a $+\pi$-kink. Middle: With a background density of 1 particle every two sites, a particle (hole) can decay into two half-particles (half-holes)
    corresponding to two consecutive particles (holes). This creates kinks of amplitude $\pm \pi/2$. Bottom: Due to their larger entropy, half-excitations are expected to be much more frequent than full-excitations.}
    \label{fig:bosonization_fermion_spin}
\end{figure}

At low energies, these models can be studied by bosonization around the half-filled density $\rho_0=1/2a$ \cite{Giamarchi,von_Delft_bosonization,Gogolin_2004,NdupuisCMUG2}. This boils down to working with two bosonic fields $\htheta(x)$ and $\hphi(x)$ that satisfy the commutation relation $[\htheta(x), \hphi(y)]=-i\frac{\pi}{2}{\rm sgn}(x-y)$. In terms of fermions, the former identifies with the fermion phase while the latter is related to the density fluctuations,
\begin{align}
    \hc_j &\sim \cos\left(\frac{\pi}{2}j -\hphi(x)\right) e^{i\htheta(x)},\\
    \label{eq:bosonized_rho}
    \frac{\hn_j}{a}&=\rho_0-\frac{1}{\pi}\nabla \hphi(x)+\frac{(-1)^j}{a\pi^2}\cos(2\hphi(x))\equiv \hrho(x).
\end{align}
with $x=ja$. In the language of spins, $e^{i\htheta(x)}$ becomes roughly the spin ladder operator $\hS^-_j$ and the field $\phi$ describes the fluctuations of the operator $\hS^z_j$. Retaining only the slowly varying component of the density, $\hrho=\rho_0-\frac{1}{\pi}\nabla \hphi$, it appears that adding a particle on top of the background density $\rho_0$, i.e., a contribution of $1$ to $\int \d x (\hrho(x)-\rho_0)$, is represented by a downward kink of amplitude $\pi$ (see Fig.~\ref{fig:bosonization_fermion_spin}, top) since
\begin{align}
    \int_0^L \d x (\hrho(x)-\rho_0)=\frac{\hphi(0)-\hphi(L)}{\pi}.
\end{align}
Because the lattice is half filled at $\rho_0$, one can also have half kinks of amplitude $- \pi/2$ corresponding to two neighbouring particles (see Fig.~\ref{fig:bosonization_fermion_spin}, middle). In general, full particles are expected to spontaneously decay into two half particles which possess a higher entropy. Similarly, additional holes are represented as kinks of height $+\pi$ and can decay into two half-kinks of height $+\pi/2$ (see Fig.~\ref{fig:bosonization_fermion_spin}, bottom). In the language of spins, full excitations are called magnons while half excitations are spinons \cite{Giamarchi}.

\begin{figure*}[t!]
    \centering
    \includegraphics[width=\linewidth]{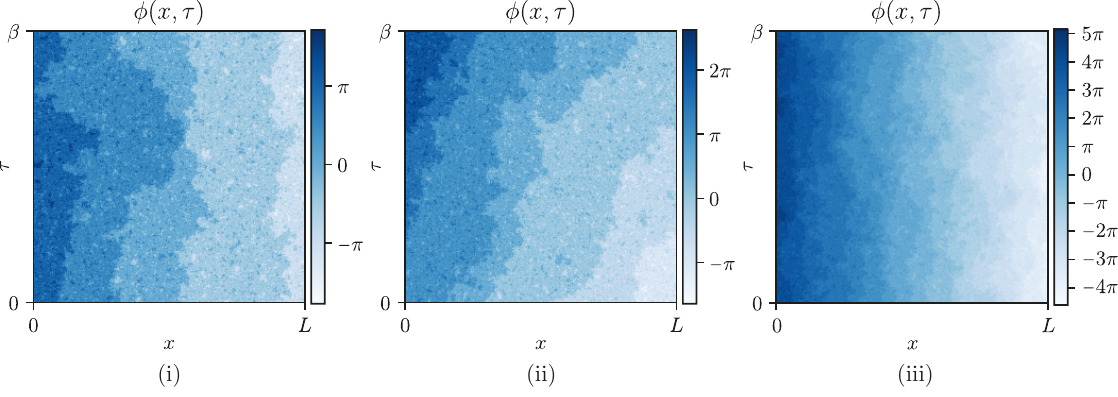}
    \caption{Typical field configurations contributing to the path integral \eqref{eq:partition_function} on a lattice of size $256 \times 256$. All kinks are of amplitude $\pi/2$. Counting the number of space and time kinks shows that figure $({\rm i})$ has $N_x=2$, $N_\tau=0$, $({\rm ii})$ has $N_x=2$, $N_\tau=-1$, and $({\rm iii})$ has $N_x=8$, $N_\tau=0$. Since kinks identify as the worldlines of half particles, $N_x$ counts the number of full particles, and $N_\tau$ encodes the average (imaginary-time) particle current $j(x,\tau)=\frac{1}{\pi}\partial_\tau \phi(x,\tau)$. The quantization of $N_\tau$ can then be seen as arising from the indistinguishability of quantum particles (kinks). Indeed, the trace in $Z={\rm Tr}\, e^{-\beta \hat{H}}$ ensures the particles return to their original position up to a permutation.}
    \label{fig:field_examples}
\end{figure*}

Using this mapping, the Hamiltonians \eqref{eq:H_fermion} and \eqref{eq:H_spin} lead to the bosonized Hamiltonian
\begin{align}\label{eq:H_bosonized}
    \hat{H}=&\int_0^L \d x \frac{u}{2 \pi}\left[ K(\nabla\htheta)^2 + \frac{1}{K}(\nabla\hphi)^2\right]\nonumber\\
    &- g \cos(4\hphi) +\frac{\mu}{\pi}\nabla \hphi,
\end{align}
where the sound velocity $u$, the Luttinger parameter $K$ and the umklapp strength $g$ can be related to the microscopic couplings through the Bethe ansatz equations \cite{Haldane_bethe}. For our purposes, it is sufficient to know that $g \sim V$. In the path-integral formalism, integrating out the field $\theta$ shows that the grand-canonical equilibrium partition function $Z={\rm Tr} \, e^{-\beta \hat{H}}$ at inverse temperature $\beta=1/T$ can be expressed as
\begin{equation}\label{eq:partition_function}
    Z=\sum_{N_x,N_\tau=-\infty}^{+\infty}\underset{\substack{\phi(0,\tau)=\phi(L,\tau) + \pi N_x \\ \phi(x,0)=\phi(x,\beta) + \pi N_\tau}}{\int \mathcal{D}\phi}\, e^{-S[\phi]},
\end{equation}
where the Euclidean action is the sine-Gordon action
\begin{align}\label{eq:phi_action}
    S[\phi]=&\int_{x,\tau} \frac{1}{2 \pi K}\left[ u(\partial_x \phi)^2+\frac{1}{u}(\partial_\tau \phi)^2\right]\nonumber\\
    &- g \cos(4\phi) +\frac{\mu}{\pi}\partial_x \phi,
\end{align}
with the notation $\int_{x,\tau}=\int_0^L \d x \int_0^\beta \d \tau$.
The boundary conditions in the path integral \eqref{eq:partition_function} are crucial to recover the correct physics. Indeed, $N_x$ identifies with the number of downwards kinks of amplitude $\pi$ encountered from $x=0$ to $x=L$ (see Fig.~\ref{fig:field_examples}). In other words, $N_x$ is the total number of particles added to the half-filled system. The quantization of $N_\tau$ is due to the fact that the field $\phi$ and $\phi+\pi$ encode the same physical state. A more formal bosonization-intensive derivation of the boundary conditions can be found in Appendix.~\ref{app:Bosonization_subtleties}.

\subsection{Lattice representations}
The field theory defined by Eqs.~\eqref{eq:partition_function} and \eqref{eq:phi_action} must be regulated by a short-distance cut-off to make sense. For numerical simulations, a convenient regulation is to put the field theory on a lattice with unit spacing in space and time. This means that the continuous field $\phi(x,\tau)$ is replaced by its discretized version $\phi_i$ such that, with $\hat{x}$ and $\hat{\tau}$ the unit vectors in the space and time directions,
\begin{equation}
    \phi_{i=i_x\hat{x}+i_\tau \hat{\tau}}\equiv \phi(x=i_x, \tau=i_\tau),
\end{equation}
where $i\in \llbracket 1,L\rrbracket \times \llbracket 1,\beta\rrbracket$ and $L$, $\beta$ are now integers. The boundary conditions are $\phi_{i + L\hat{x}}=\phi_i - \pi N_x$ and $\phi_{i+\beta\hat{\tau}}=\phi_i - \pi N_\tau$, and the action \eqref{eq:phi_action} becomes
\begin{align}\label{eq:S_discretized}
    S(\phi)\equiv\sum_i &\frac{1}{2\pi K}\left[ (\phi_i-\phi_{i+\hat{x}} )^2 + (\phi_i-\phi_{i+\hat{\tau}} )^2\right]\nonumber\\
    &- g\cos(4\phi_i) + \frac{\mu}{\pi}(\phi_{i+\hat{x}}-\phi_i),
\end{align}
where we have set $u=1$ for simplicity. This is the \emph{bosonic} representation of the lattice field theory.

Splitting the field $\phi$ into its winding numbers $N_x$, $N_\tau$ and a periodic field $\varphi_i = \phi_i+\pi N_x\frac{i_x}{L} + \pi N_\tau \frac{i_\tau}{\beta}$ yields the \emph{winding} representation of the action
\begin{align}\label{eq:S_winding}
    S(N_x,&N_\tau,\varphi)\equiv\frac{\pi \beta L}{2K}\left[\left( \frac{N_x}{L} \right)^2 + \left( \frac{N_\tau}{\beta} \right)^2 \right]-\mu \beta N_x\nonumber\\
    &+\sum_i \frac{1}{2\pi K}\left[ (\varphi_i-\varphi_{i+\hat{x}} )^2 + (\varphi_i-\varphi_{i+\hat{\tau}} )^2\right]\nonumber\\
    &- g\cos\left(4\varphi_i - 4\pi N_x\frac{i_x}{L}-4\pi N_\tau\frac{i_\tau}{\beta}\right),
\end{align}
which shows that the windings and $\varphi$ are only coupled through the cosine term. This winding representation is mainly exploited in Sec.~\ref{sec:observables} to discuss the physics of the model.

\begin{figure}[h!]
    \centering
    \includegraphics[width=0.65\linewidth]{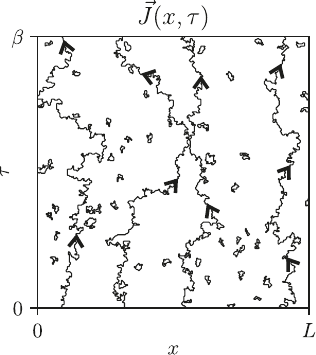}
    \caption{Current field $\vec{J}$ obtained from the field depicted in Fig.~\ref{fig:field_examples}-$({\rm i })$ by using Eqs.~\eqref{eq:n_f_def} and \eqref{eq:current_def}. The current traces out the (oriented) topographic lines of the field $\phi$. For the sake of clarity, we do not display the very small current loops that would otherwise cover up the picture.}
    \label{fig:loop_field}
\end{figure}

\subsection{Current-fluctuation representation}
We now present a \emph{current-fluctuation} representation of the field theory in terms of loops formed by the coarse grained particle current, and small fluctuations of the density. The particle current is obtained by drawing the topographic lines (or contour lines) of the field $\phi$ (see Fig.~\ref{fig:loop_field}), a procedure akin to the Coulomb gas formalism \cite{Nienhuis_1982,Jacobsen_2009} used in conformal field theory. This representation will serve as the starting point of the Monte Carlo algorithm developed in Section~\ref{sec:MC_algorithm}.

To be more precise, we first make apparent the discrete height map embedded in the field $\phi$ by writing
\begin{align}\label{eq:n_f_def}
    \phi_i \equiv \frac{\pi}{2} (n_i + f_i),
\end{align}
with $n_i \in \mathbb{Z}$ the discrete height field, and $f_i\in ]-1/2,1/2]$ the fluctuation field. The field $f_i$ is periodic in all directions while the height map $n_i$ obeys the boundary conditions $n_{i+L\hat{x}} = n_i - 2 N_x$ and $n_{i+\beta\hat{\tau}} = n_i - 2 N_\tau$. With these variables, the action \eqref{eq:S_discretized} becomes
\begin{align}
    S(n,f)=\sum_i& \frac{\pi}{8 K} \Big[(n_i - n_{i+\hat{x}}+f_i - f_{i+\hat{x}})^2\nonumber\\
    &\hspace{0.5cm} + (n_i - n_{i+\hat{\tau}} + f_i - f_{i+\hat{\tau}} )^2 \Big] \nonumber\\
    &- g \cos(2\pi f_i)+ \frac{\mu}{2} (n_{i+\hat{x}}- n_i).
\end{align}

\begin{figure}[h!]
    \centering
    \includegraphics[width=0.8\linewidth]{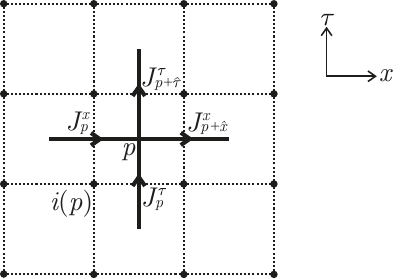}
    \caption{The fields $\phi_i$, $n_i$, $f_i$ are all defined on the 2D lattice drawn in dashed lines. Its sites are labelled by $i$ and its plaquettes by $p$. The current field $\vec{J}_p$ (in solid lines) introduced in \eqref{eq:current_def} lives on the edges of the dual lattice.}
    \label{fig:def_lattice_objects}
\end{figure}
We now introduce the two-component current field $\vec{J}_p=(J^x_p,J^\tau_p)$. It connects plaquettes $p$ of the lattice --- hence the subscripts in $J_p^{x/\tau}$ --- and thus lives on the edges of the dual lattice (see Fig.~\ref{fig:def_lattice_objects}). It is defined by
\begin{align}\label{eq:current_def}
    J^x_p \equiv n_{i(p)+\hat{\tau}} - n_{i(p)},\quad J^\tau_p \equiv n_{i(p)}-n_{i(p)+\hat{x}},
\end{align}
where $i(p)=p-(\hat{x}+\hat{\tau})/2$ connects the coordinates of the two lattices. From the previous definition, it follows that the current field is periodic in all directions and also divergenceless since
\begin{align}
    (\vec{\nabla} \cdot \vec{J})_p \equiv J^x_{p+\hat{x}} - J^x_p + J^\tau_{p+\hat{\tau}} - J^\tau_p = 0,
\end{align}
which implies that the current is conserved and thus forms loops. Using this mapping, one arrives at the \emph{current-fluctuation} representation of the action
\begin{align}\label{eq:current_fluc_action}
    S(\vec{J},f)\equiv\sum_p &\frac{\pi}{8K} \Big[(J^\tau_p+ f_{i(p)} - f_{i(p)+\hat{x}} )^2\nonumber\\
    &+ (J^x_p + f_{i(p)+\hat{\tau}} - f_{i(p)})^2 \Big]\nonumber\\
    &- g \cos(2\pi f_{i(p)} ) - \frac{\mu}{2} J^\tau_p.
\end{align}
Note that if one neglects the fluctuations by setting $f_i=0$, one recovers a $(1+1)$D version of the $(2+1)$D link-current representation of the quantum rotor model \cite{Wallin_1994_2DSI}. The link-current representation is obtained from the Bose-Hubbard model by using a Villain approximation and integrating out the phase fluctuations, while the current-fluctuation representation we propose relies on bosonization to decouple the density and the phase modes, before integrating out the phase fluctuations. Denoting by $\mathcal{C}_0$ the set of divergenceless current fields, the partition function is
\begin{align}
    Z=\sum_{\vec{J}\in \mathcal{C}_0} \prod_i \int_{-\frac12}^{\frac12} \d f_i \, e^{-S(\vec{J},f)}.
\end{align}
Since the bosonic field $\phi$ has the same physical meaning as $\phi+\pi$, the mapping from $\phi$ to $(\vec{J},f)$ can be inverted by first requiring that $n_{i=0}=0$, then using Eq.~\eqref{eq:current_def} to recover the discrete height field $n_i$, and finally using Eq.~\eqref{eq:n_f_def} to get $\phi_i$.

\section{Monte Carlo algorithm}
\label{sec:MC_algorithm}

\subsection{The need for an efficient sampling algorithm}
\label{sec:need_for_efficient_algo}
A standard Monte Carlo algorithm to sample the probability distribution $\pi(\phi)\equiv e^{-S(\phi)}/Z$ is the Metropolis--Hastings algorithm \cite{Metropolis1953} where one proposes an update $\phi \to \phi'$ and accepts it with probability
\begin{equation}
    p_{\rm Met.}=\exp(-[S(\phi')-S(\phi)]_+),
\end{equation}
with $[x]_+=\max(0,x)$. In order to be ergodic, the algorithm has to propose local updates $\phi_i \o \phi_i + \delta$ and global updates which change the topological numbers $N_x$, $N_\tau$. Such a possible straightforward update is to change $N_x\to N_x +1$ (respectively $N_\tau \to N_\tau+1$) while keeping $\phi$ constant, which amounts to creating a kink of amplitude $\pi$ at the boundary $x=L$ (respectively $\tau=\beta$). This implies an increase of the quadratic part of the action at the boundary of the order of $\beta \pi/2 K$ (respectively $L \pi/2 K$). For $K=0.35$ and $\mu\in [0,1]$ (which we use in all our following simulations), the variation of the total action is dominated by the quadratic contribution, so one finds $p_{\rm Met.}\simeq \exp(-\pi/(2\times0.35))^\beta=(1.12\times 10^{-2})^\beta$ (and similarly for $\beta \leftrightarrow L$). This is a ridiculously small acceptance rate which, along with its exponential scaling, should discourage anyone from trying to attempt large system-size simulations with the Metropolis--Hastings algorithm. We also expect cluster algorithms \cite{Swendsen_1987,Wolff_cluster_algo} to be very inefficient for sampling the sine-Gordon model. Indeed, cluster algorithms used for bosonized systems \cite{Hasenbuch1994_clusterSG,Werner2005clusterJJ,bouvdup_2025_bosonized1d} cannot change the boundary conditions, and, furthermore, crucially rely on the particle-hole symmetry $\phi \to - \phi$ which is broken by the chemical potential $\mu$.

To overcome this issue, we propose a \emph{Worm algorithm} (Wo) and its enhanced version, the \emph{Smooth Worm algorithm} (SmoWo), which perform two types of updates:
\begin{itemize}
    \item local $\phi$ updates at constant $N_x$ and $N_\tau$ using the Event-Chain Monte Carlo (ECMC) algorithm,
    \item $N_x$ and $N_\tau$ updates using the worm algorithm, enhanced with ECMC moves for the SmoWo algorithm.
\end{itemize}
In a nutshell, ECMC algorithms \cite{Michel2014GenECMC,krauth2021ECMC} are a class of continuous-time and rejection-free algorithms that are non-reversible as they do not satisfy detailed balance but only the weaker global balance.

They were first introduced as a non-reversible extension of the Metropolis algorithm, and formally realise a piecewise deterministic Markov process \cite{Monemvassitis_2023}. They can be used for any system with continuous degrees of freedom and have been shown to perform far better than classical reversible schemes such as the Metropolis--Hastings algorithm in various contexts \cite{Nishikawa_2015,Michel2015spin,Klement_2021,bouvdup_2025_bosonized1d}. The worm algorithm \cite{Prokofiev_1998_worm,Prokofiev_2001_worm} can be applied to any model whose configurations are made of loops, which for the sine-Gordon model we have identified in the current-fluctuation representation \eqref{eq:current_fluc_action}. By allowing intermediate configurations to have one open path, the ``worm", the worm algorithm drastically reduces the critical slowing down observed near phase transitions. In the following, we will consider a worm made of two   with multiple paths, as 

In the following, we first extend the \emph{current-fluctuation} representation \eqref{eq:current_fluc_action} to include a worm carrying two half-particles (see also \cite{Kuklov_2004,Soyler_2009} for other uses of multiple worms). We then describe the algorithms in detail and test their performance. Technical details such as a proof that both algorithms satisfy the global balance condition can be found in Appendix.~\ref{app:code}.

\begin{figure}[h!]
    \centering
    \includegraphics[width=0.85\linewidth]{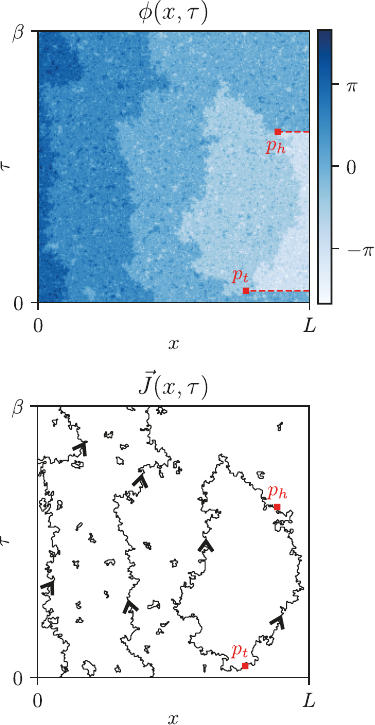}
    \caption{Top: A field configuration $\phi$ with a worm ranging from $p_t$ to $p_h$ and obtained for a system of size $L=\beta=256$. The red lines are artificial discontinuities of amplitude $\pi$ which are needed to represent the field $\phi$. Bottom: Current field $\vec{J}$ associated to $\phi$. It is divergenceless everywhere except at $p_h$ and $p_t$. For the sake of clarity, we do not display the very small current loops that would otherwise cover up the picture.}
    \label{fig:loop_worm}
\end{figure}

\subsection{Extended current-fluctuation representation}
To implement any worm algorithm, we must extend the loops formed by the current field $\vec{J}$ in the current-fluctuation representation \eqref{eq:current_fluc_action} by allowing for an open path called the worm. However, instead of attaching a single current line between the worm's head and tail, we attach two lines (see Fig.~\ref{fig:loop_worm}). This will enable the computation of the phase-phase correlation function in Sec.~\ref{sec:theta_2point_function}. Calling $\mathcal{C}_2(p_t,p_h)$ the set of divergenceless current fields except at $p_t$ (the worm's tail) where $(\vec{\nabla} \cdot \vec{J})_{p_t}=2$, and at $p_h$ (the worm's head) where $(\vec{\nabla} \cdot \vec{J})_{p_h}=-2$, we therefore define the extended model by its partition function
\begin{align}\label{eq:MC_partition_function}
    Z_{\rm w}=\sum_{p_t,p_h} \sum_{\vec{J}\in \mathcal{C}_2(p_t,p_h)} \prod_i \int_{-\frac12}^{\frac12} \d f_i \, e^{-S(\vec{J},f)}.
\end{align}
In terms of this extended partition function, the physical partition function $Z$ reads
\begin{equation}\label{eq:Z_bf}
    Z=Z_{\rm w} \langle \delta_{p_t,0}\delta_{p_h,0} \rangle_{\rm w}= \frac{Z_{\rm w}}{\beta L}\langle \delta_{p_t,p_h} \rangle_{\rm w},
\end{equation}
where $\langle \bullet \rangle_{\rm w}$ is the average with respect to the model defined in \eqref{eq:MC_partition_function} and the second equality comes from the translational invariance of the system.

\subsection{Worm algorithm (Wo)}
\label{sec:Wo_algorithm}
The Wo algorithm consists of two types of moves: the ECMC moves, and the worm moves.

\paragraph{ECMC moves.} We first implement some ECMC moves to deal with the fluctuations $f$. To study the configuration space $\Omega=\{(\vec{J},f)\}$, an ECMC algorithm works in the augmented space $\Omega \times V$ with $V=\{ (i \in  \llbracket 1, L\rrbracket\times \llbracket 1, \beta \rrbracket, e \in \{-1,1\}) \}$. The elements $v=(i,e)\in V$ are called \emph{lifting} variables. As long as $v=(i,e)$ is not updated, we continuously increase $f_i$ if $e=+1$ and decrease it if $e=-1$. The exact form of this deterministic motion is
\begin{align}\label{eq:ecmc_motion}
    \partial_t f_j(t)=\begin{cases}
    e \text{ if } j=i,\\
    0  \text{ otherwise},
\end{cases}
\end{align}
where $t$ is the time of the Markov process. In principle, when $f_i$ reaches $\pm 1/2$, its neighbouring currents should be updated so as to continue the motion from $f_i = \mp 1/2$ and keep $f_i \in [-1/2,1/2]$. However, for the sake of simplicity, we decide not to do so since it does not affect the reconstructed field $\phi(\vec{J},f)$ which caries the physical information. The evolution \eqref{eq:ecmc_motion} proceeds until a random event updates $v\to v'$, and then resumes with $v'$. As is usually done in ECMC algorithms, we associate an event with each interaction in the model. In the current-fluctuation representation \eqref{eq:current_fluc_action}, we identify quadratic interactions like
$S_\q^{i,i+\hat{x}}=\frac{\pi}{8 K}(J^\tau_{p(i)}+f_i-f_{i+\hat{x}})^2$, and cosine interactions $S_\c^i=-g\cos(2\pi f_i)$. The lifting variable $v=(i,e)$ may trigger 4 possible quadratic events stemming from the interactions $S_\q^{i,j}$ with $j=i\pm \hat{x},i\pm \hat{\tau}$ and one cosine event linked to $S_\c^i$. The associated time-dependent event rates are
\begin{align}
    \lambda^{i,j}_\q(t)&=[e\partial_{f_i(t)} S_\q^{i,j}(t)]_+,\\
    \lambda^i_\c(t)&=[e\partial_{f_i(t)} S_\c^i(t)]_+.
\end{align}
Intuitively, these events are triggered when the deterministic motion stores too much energy (or action) in an interaction. Using the pairwise symmetries of the interactions, an energy excess in $S_\q^{i,j}$ obtained from $v=(i,e)$ can be released by setting $v'=(j,e)$. Similarly, an excess in $S_\c^i$ with $v=(i,e)$ can be mitigated by setting $v'=(i,-e)$ to undo the previous move. We therefore decide that the rate $\lambda_\q^{i,j}$ triggers the update $(i,e) \to (j,e)$ while $\lambda_\c^i$ triggers $(i,e) \to (i,-e)$.
To ensure ergodicity, an additional \emph{refreshment} event is customarily added. It has a constant (i.e. configuration-independent) rate $\lambda_{\rm r}$ and uniformly draws a new lifting variable $v'\in V$. Anticipating the worm moves presented below, we also introduce a constant rate $\lambda_{\rm w}$ for triggering such a move. 

In practice, the ECMC algorithm is simulated by i) finding which event occurs first and at what event time, ii) updating the fluctuation field $f$ to that time using \eqref{eq:ecmc_motion}, iii) performing the update $v \to v'$ triggered by the event and iv) repeating the procedure. For the algorithm described above, there are four types of event times to compute when moving $f_i$ along $e$: the quadratic ones $t_\q^{i,j}$, the cosine one $t_\c^i$, the refreshment one $t_{\rm r}$ and the worm one $t_{\rm w}$. These event times can be explicitly computed using inversion sampling (see Appendix~\ref{app:event_times}).

\begin{figure}
    \centering
    \includegraphics[width=\linewidth]{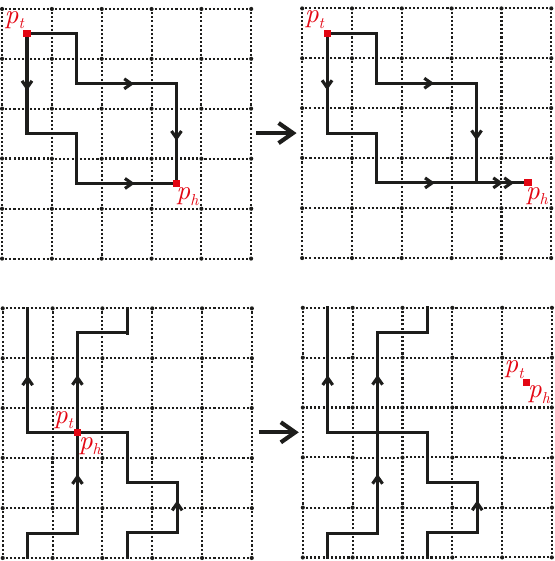}
    \caption{The worm can perform two type of updates. (Top) The \emph{shift} update shifts the head $p_h$ of the worm to a neighbouring plaquette $p_h'$ and changes the current along its path by $2$ if $p_h' \in \{ p_h+\hat{x}, p_h + \hat{\tau} \}$ and by $-2$ if $p_h' \in \{ p_h-\hat{x}, p_h - \hat{\tau} \}$. (Bottom) The \emph{move} update can only happen when $p_h=p_t$ and moves both points to a randomly picked new location $p_h'=p_t'$.}
    \label{fig:worm_explaination}
\end{figure}

\paragraph{Worm moves.} In the general ECMC framework introduced in the above, we have added a rate $\lambda_{\rm w}$ at which worm moves occur. We now describe these moves which update the current field $\vec{J}$. We consider two types of worm updates: the \emph{shift} update (see Fig.~\ref{fig:worm_explaination}, top) and the \emph{move} update (see Fig.~\ref{fig:worm_explaination}, bottom). If the worm's head has reached its tail ($p_h=p_t$), with probability $1/2$ we propose a move update which moves both $p_h$ and $p_t$ to a new common location $p_h'=p_t'$ uniformly chosen on the dual lattice. Otherwise, we propose a shift update which randomly selects a neighbour $p_h' \in \{p_h \pm \hat{x},p_h \pm \hat{\tau}\}$ and shifts $p_h$ to $p_h'$. If $p_h \ne p_t$, we always propose a shift update. The move update does not change the action of the configuration, so it is always accepted to obey detailed balance. The shift update is uniformly picked among the four possible directions and is accepted with the Metropolis--Hastings filter
\begin{equation}\label{eq:worm_p_met}
    P(p_h\to p_h')=\min(1,R e^{-\Delta S}),
\end{equation}
with
\begin{equation}
    R=\begin{cases}
         2 &\text{ if } p_h=p_t,\\
         1/2 &\text{ if }p_h'=p_t,\\
         1 &\text{ otherwise,}
    \end{cases}
\end{equation}
where $\Delta S$ is the variation of the action \eqref{eq:current_fluc_action}. This ends the definition of the Wo algorithm, a pseudocode implementation of which can be found in Appendix~\ref{app:pseudo_code}.

\begin{figure}[h!]
    \centering
    \includegraphics[width=\linewidth]{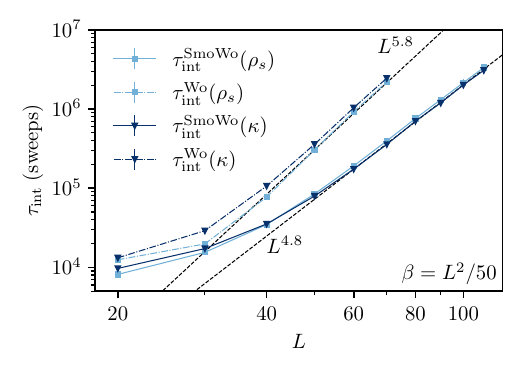}
    \includegraphics[width=\linewidth]{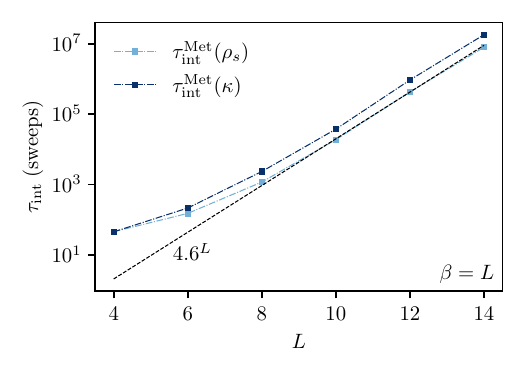}
    \caption{Integrated autocorrelation times $\tau_{\rm int}$ for the compressibility $\kappa$ and the superfluid stiffness $\rho_s$ with $g=1$, $K=0.35$, $\mu=0.32$ which corresponds to the critical region. (Top) Results for the Wo and SmoWo algorithms. The system sizes are scaled as $\beta = L^2/50$ since the dynamical exponent of the transition is $z=2$ (see Sec.~\ref{sec:numerical_results}), and we took $\lambda_{\rm w}=1$, $\lambda_{\rm r}=0.1/(L\beta)$. This leads to identifying the algorithmic dynamical exponents $z_{\rm alg.}\simeq 5.8$ for the Wo algorithm, and $z_{\rm alg.}\simeq 4.8$ for the SmoWo algorithm. (Bottom) Results for the Metropolis--Hastings (Met) algorithm discussed in Sec.~\ref{sec:need_for_efficient_algo}. We take $\beta=L$ to accommodate the very small system sizes. The integrated autocorrelation time scales exponentially as $\tau_{\rm int}\sim 4.6^L$.}
    \label{fig:t_int}
\end{figure}

\paragraph{Performance} To test the Wo algorithm described previously, we define the algorithmic time $\texttt{t}$ expressed in sweeps (i.e. $\beta L$ operations) as increasing by $1/(\beta L)$ each time an event time or a worm acceptance rate \eqref{eq:worm_p_met} is computed. The algorithmic time is expected to scale as the CPU time while being less sensitive to specific code implementation details. We then consider the autocorrelation function defined for an observable $\mathcal{O}$ as
\begin{align}\label{eq:def_autocorr}
    C_\mathcal{O}(\texttt{t})=\frac{\langle \mathcal{O}(\texttt{t}) \mathcal{O}(0)\rangle -\langle \mathcal{O} \rangle^2}{\langle \mathcal{O}^2\rangle -\langle \mathcal{O} \rangle^2}.
\end{align}
We extract from it the integrated autocorrelation time $\tau_{\rm int}^\mathcal{O}=\frac{1}{2}\sum_{\texttt{t}=-\infty}^{+\infty}C_\mathcal{O}(\texttt{t})$ which is the time needed to generate a new independent sample once the Markov process has thermalized (or mixed \cite{Levin_2017}). To assess the critical slowing-down of the algorithm, we perform this analysis at the critical point of the generic Mott transition by scaling $\beta \sim L^2$ since the dynamical critical exponent is $z=2$ (see Sec.~\ref{sec:numerical_results}). We also take $\lambda_{\rm w}=1$ which was numerically found to be optimal (the optimal region is actually quite large, roughly $\lambda_{\rm w} \in [0.5,2]$) and take $\lambda_{\rm r}=0.1/(\beta L)$ to allow ECMC moves to be correlated on the scale of the system. The results for the compressibility $\kappa$ and superfluid stiffness $\rho_s$ are shown in Fig.~\ref{fig:t_int} (top). These observables encode the large-scale fluctuations of the field $\phi$ in the space and time directions (their exact definitions can be found in Sec.~\ref{sec:scalar_observables}). It appears that for both observables $\tau_{\rm int}\sim L^{z_{\rm alg.}}$ with $z_{\rm alg.} \simeq 5.8$ the algorithmic dynamical exponent (not to be confused with the dynamical critical exponent $z$). Although this value of $z_{\rm alg.}$ is quite high, the polynomial scaling of this algorithm is a dramatic improvement over the exponential scaling (see Fig.~\ref{fig:t_int} (bottom)) of the naive Metropolis algorithm described in Sec.~\ref{sec:need_for_efficient_algo}. The difference between both scalings is traced back to the fact that the worm algorithm can reach many configurations and therefore select a typical one, while the rigid Metropolis--Hastings updates only propose very atypical configurations.

The Wo algorithm can nevertheless be improved by noticing that is suffers from a backtracking problem: the current worm moves want to create sharp kinks (i.e., with a narrow width) in the field $\phi$ which are atypical. This leads to a high rejection rate unless the attempted worm move undoes the previous one. Such backtracking behaviour is typical of simple worm algorithms and is usually minimized by using \emph{directed} worm algorithms \cite{Syljuasen_2002_directed_loop} which locally minimize the backtracking (or bounce) probability. However, for the sine-Gordon model, we found that these algorithms do not provide any significant performance gain. We also tried, with no greater success, other worm-like algorithms such as the geometrical worm algorithm \cite{Alet_2003_geometricWA} and its directed version \cite{Alet_2003_directedWA,Alet_2004_CI} which leverage properties of Markov chains with absorbing states \cite{Novotny_1995_AMC,Bortz_1975_Nfold} to grow an entire worm in a single worm event. The failure of these algorithms comes from the fact that they do not tackle the ``sharp kink" issue discussed above.

\subsection{Smooth worm algorithm (SmoWo)}
In order to solve the ``sharp kink" issue, we propose the SmoWo algorithm which, compared to the Wo algorithm, performs additional ECMC moves around the worm's head to smooth it out while it is being built. To implement these new moves, we must add a new lifting variable $\nu=(\alpha,\varepsilon)\in \mathcal{V}$ (we use Greek letters for the new lifting variable and Roman letters $v=(i,e)$ for the old). The site $\alpha$ should always be close to the worm's head $p_h$ to ensure we are smoothing around it. In practice, we choose for $\alpha$ to always be one of the four closest sites to $p_h$, that is to say
\begin{equation}
    \alpha \in \partial p_h = \left\{p_h \pm \frac{\hat{x}}{2}  \pm \frac{\hat{\tau}}{2} \right\}.
\end{equation}
To decide which of the lifting variables is being used for the deterministic motion \eqref{eq:ecmc_motion}, we introduce the variable $\sigma$ which is $0$ when using $v$ and $1$ when using $\nu$. If $\sigma=0$, the ECMC events include the refreshment event (which only updates $v$), the quadratic and cosine events and the worm update. The only event that we modify is the worm update. It is still triggered by a constant rate $\lambda_{\rm r}$ but now consists in first setting $\sigma=1$, then attempting a worm move, and finally refreshing (resampling uniformly) $\nu \in \partial p_h \times \{ -1, 1\}$ only if the worm move has been accepted (to ensure that $\alpha$ stays in $\partial p_h$). If $\sigma=1$, we define three event types: the quadratic events, the cosine events and the $\sigma$-update (there is no need for a refreshment update since $\nu$ can already be refreshed during the worm moves). When a quadratic event occurs with a neighbouring site $\gamma$, the lifting variable is updated to
\begin{align}(\alpha,\varepsilon) \to
\begin{cases}
    (\gamma,\varepsilon) &\text{ if } \gamma \in \partial p_h,\\
    (\alpha,-\varepsilon) &\text{ otherwise},
\end{cases}
\end{align}
so as to satisfy the condition $\alpha \in \partial p_h$. For a cosine neighbour, we always perform
\begin{align}
(\alpha,\varepsilon) \to (\alpha,-\varepsilon).
\end{align}
Finally, the new $\sigma$ update is triggered by the constant rate $\lambda_{\rm w}$ and simply sets $\sigma=0$. A proof of the validity of the algorithm and a pseudocode implementation are given respectively in Appendixes~\ref{app:code} and \ref{app:pseudo_code}.

Performance wise, the integrated autocorrelation time of the SmoWo algorithm is given in Fig.~\ref{fig:t_int} for the compressibility $\kappa$ and the superfluid stiffness $\rho_s$. The net result is a decrease of the algorithmic dynamical critical exponent from $z_{\rm alg.}\simeq 5.8$ for the Wo algorithm to $z_{\rm alg.}\simeq 4.8$ for the SmoWo algorithm. As shown in the next sections, this change is crucial as the finite-size effects of the model are very large and require going to sizes $L\ge 80$ to analyse the phase transition, something not possible with the Wo algorithm. The exponent $z_{\rm alg.} \simeq 4.8$ --- which could also be stated as $\tau_{\rm int} \sim \beta^{2.4}$ since we scale $\beta \propto L^2$ --- may still seem large since one usually expects a very low exponent $z_{\rm alg.}$ for worm algorithms \cite{Prokofiev_2001_worm}. A possible explanation for the remaining critical slowing down may lie in the slow separation dynamics of the worm of height $\pi$ into two kinks of height $\pi/2$. We have also tried larger smoothing areas $\partial p_h$ with no notable performance change.

\section{Observables and physical insights}
\label{sec:observables}
\subsection{Phase diagram}
\begin{figure}[h!]
    \centering
    \includegraphics[width=0.8\linewidth]{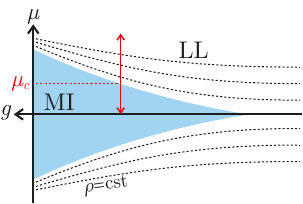}
    \caption{Phase diagram of the Mott transition. As $g$ is increased, the Mott insulator (MI) lobe expands at the expense of the Luttinger liquid (LL). The dashed lines are lines of constant density and the MI is at the fixed density $\rho_0=\frac{1}{2a}$. In this work, we focus on the \emph{generic} transition which is crossed by varying $\mu$ as shown by the red arrow. The transition at the tip of the lobe is different as it belongs to the BKT (Berezinskii--Kosterlitz--Thouless) universality class \cite{Giamarchi_1997_Mott1D,Fisher_1989_SI,Daviet_2019_sG}.}
    \label{fig:Mott_lobe}
\end{figure}

The sine-Gordon action \eqref{eq:phi_action} can be used to get a intuitive understanding of the Mott transition. On the one hand, the chemical potential $\mu$ wants to tilt the field $\phi$ and create many kinks, while on the other hand, a strong coupling $g$ penalizes their creation. In the limit of negligible $g$, the winding representation \eqref{eq:S_winding} of the action becomes
\begin{align}\label{eq:LL_action}
    S_{\rm LL}&=\frac{\pi \beta L}{2K}\left[u\left( \frac{N_x}{L} - \frac{K\mu}{\pi u}\right)^2 + \frac1u\left( \frac{N_\tau}{\beta} \right)^2 \right] \\
    &+\sum_i \frac{1}{2\pi K}\left[ u(\varphi_i-\varphi_{i+\hat{x}} )^2 + \frac{1}{u}(\varphi_i-\varphi_{i+\hat{\tau}} )^2\right],\nonumber
\end{align}
where we have explicitly reintroduced the speed $u$. This shows that the topological variables $N_x$, $N_\tau$ decouple from the periodic field $\varphi$. The action for $\varphi$ is that of a Luttinger liquid (LL). In the opposite limit of very large $g$, all kinks are suppressed so $N_x=N_\tau=0$, and the field $\varphi$ gets pinned around a minimum of $\cos(4\varphi)$, e.g. $\varphi=0$. One can thus expand the cosine to obtain
\begin{align}\label{eq:MI_action}
    S_{\rm MI}=\sum_i & \frac{1}{2\pi K}\left[ u(\varphi_i-\varphi_{i+\hat{x}} )^2 + \frac{1}{u}(\varphi_i-\varphi_{i+\hat{\tau}} )^2\right]\nonumber\\
    &+8g \varphi_i^2.
\end{align}
This is the Mott insulator (MI). We thus expect the existence of two critical points $\pm \mu_c$ (see Fig.~\ref{fig:Mott_lobe}) such that the system is described by the LL action \eqref{eq:LL_action} for $|\mu|>\mu_c$ (with potentially renormalized couplings $K,u,\mu \to K_\R,u_\R,\mu_\R$), and by the MI action \eqref{eq:MI_action} for $|\mu|<\mu_c$ (with, again, $K,u,g\to K_\R,u_\R,g_\R$). In the following, we focus on the transition at $+\mu_c>0$ without loss of generality since the action \eqref{eq:phi_action} is invariant under $\mu,\phi \to -\mu, -\phi$.

\subsection{Scalar observables}
\label{sec:scalar_observables} 
An order parameter for the transition is the doping $\delta \rho=\langle \hrho \rangle - \rho_0 $ (or the magnetization $m=\langle \hS_j^z\rangle$ for spins) which, in terms of the bosonized variables, is given by
\begin{align}\label{eq:delta_rho}
    \delta \rho \equiv \left\langle \frac{N_x}{L} \right\rangle.
\end{align}
Since there are no kinks in the MI, $\delta \rho$ vanishes. In the LL, in the limit of $L,\beta \to \infty$, it appears from Eq.~\eqref{eq:LL_action} that the doping concentrates around $\frac{K_\R \mu_\R}{\pi u_\R}$.

Another order parameter can be found by considering the rapid fluctuations of the density $(-1)^{i_x}\cos(2\phi_{i=i_x\hat{x}+i_\tau \hat{\tau}})$ (see Eq.~\eqref{eq:bosonized_rho}). Using the periodic field $\varphi_i=\phi_i+\pi N_x i_x/L+\pi N_\tau i_\tau/\beta$, these fluctuations can be rewritten as
\begin{equation}
    \cos\left(2\varphi_i -2\pi\left(\frac{1}{2} + \frac{N_x}{L}\right) i_x -2\pi \frac{N_\tau}{\beta} i_\tau \right),
\end{equation}
and thus appear as fluctuating in space at twice the Fermi wave-vector $k_F\equiv\pi(\rho_0+ \delta \rho)=\pi (1/2+N_x/L)$. Neglecting for simplicity $N_\tau$, the amplitude of this $2k_F$ modulation is
\begin{align}\label{eq:C_2kF}
    C_{2k_F}\equiv\langle \cos(2\varphi_i -2 \overline{\varphi})\rangle,
\end{align}
where we have removed the average $\overline{\varphi}=(\beta L)^{-1}\int_{x,\tau} \varphi$ which just fixes the overall sign of the density modulation \footnote{This is analogous to what is done for the Ising model. Instead of computing the magnetization $m$, one usually considers its absolute value $|m|$ \cite{Newman_MC_StatPhys}.}. In the LL, one can show using the action \eqref{eq:LL_action} that $C_{2k_F}$ should vanish. In the MI, $C_{2k_F}>0$ since $\varphi_i-\bar \varphi$ gets locked around $0$.

The Mott transition is also characterized by the compressibility $\kappa$ and the superfluid stiffness $\rho_s$ which respectively measure the response of the system to a shift in the spatial boundary conditions (i.e., a change of the density) and in the temporal boundary conditions (i.e., a change of the particle-current). In terms of the bosonic variables, these observables are defined as
\begin{align}\label{eq:kappa_def}
    \kappa &\equiv \lim_{q\to 0}\frac{q^2}{\pi^2}\langle |\varphi(q,0)|^2 \rangle, \\
    \label{eq:rho_s_def}
    \rho_s &\equiv \lim_{\omega_n \to 0}\frac{\omega_n^2}{\pi^2}\langle |\varphi(0,\omega_n)|^2 \rangle,
\end{align}
with $\varphi(q,\omega_n)=(\beta L)^{-1/2}\int_{x,\tau} e^{-i(qx+\omega_n \tau)}\varphi(x,\tau)$ (see Appendix.~\ref{app:kappa_rho_s} for additional comments on these definitions). For finite-size systems, the limit $q \to 0$ (respectively $\omega_n \to 0$) is approached by using the smallest possible momentum, $\frac{2\pi}{L}$ (respectively frequency, $\frac{2\pi}{\beta}$). From \eqref{eq:LL_action}, it is straightforward to show that $\kappa = \frac{K_\R}{\pi u_\R}$ and $\rho_s=K_R u_\R/\pi$ from which one can extract $K_\R=\pi \sqrt{\kappa \rho_s}$ and $u_\R=\sqrt{\rho_s/\kappa}$. For the MI action \eqref{eq:MI_action}, one instead gets $\kappa=\rho_s=0$.

The last $\varphi$-dependant observable we consider is the two-point function
\begin{align}
    C_\varphi(i)\equiv \langle e^{i(\varphi_i-\varphi_{i=0})} \rangle,
\end{align}
which, from \eqref{eq:LL_action} and \eqref{eq:MI_action}, decays algebraically in the LL as $C_\varphi(i_x\hat{x})\sim i_x^{-2K_\R}$ (and similarly in time), and reaches a plateau in the MI because $\varphi$ gets locked around a minimum. Physically, this two-point function is akin to $\langle e^{i2(\varphi_i - \varphi_{i=0})} \rangle$ which is the $2k_F$ part of the density-density correlation function.

The field $\phi$ can only be rebuilt from $(\vec{J},f)$ when $p_h=p_t$, i.e., when there is no worm. To compute an observable $\mathcal{O}(\phi)$, we thus check at fixed time intervals $T_{\rm sample}$ if $p_h=p_t$, and if so, we output a sample $\mathcal{O}(\phi(\vec{J},f))$. The statistical average $\langle \mathcal{O}(\phi) \rangle$ is then given by the sample average. A pseudocode implementation of the computation of observables can be found in Appendix~\ref{app:pseudo_code}.

\subsection{Phase-phase two-point function}
\label{sec:theta_2point_function}
We now focus on the phase-phase two-point function $C_\theta(x_h-x_t,\tau_h-\tau_t)=\big\langle e^{i(\theta(x_h,\tau_h)-\theta(x_t,\tau_t))} \big\rangle$. At the operator level, it is defined as
\begin{align}\label{eq:two_point_def}
    &C_\theta(x_h - x_t, \tau_h - \tau_t) \nonumber\\
    & \equiv \frac{ {\rm Tr}\left( \hat{\mathcal{T}}e^{\tau_h\hat{H}} e^{i\htheta(x_h)} e^{(\tau_t-\tau_h)\hat{H}} e^{-i\htheta(x_t)} e^{-\tau_t\hat{H}} e^{-\beta \hat{H}} \right)  }{Z},
\end{align}
with $\hat{\mathcal{T}}$ the (imaginary) time-ordering operator. For bosons, $\hc_j \sim e^{i\htheta(x)}$ and this two-point function corresponds to creating a particle at $(x_t,\tau_t)$ and destroying it at $(x_h,\tau_h)$. Since spins can be mapped onto hardcore bosons, one similarly finds that $S^-_j \sim (-1)^j e^{i\theta(x)}$. However, for fermions, $\hc_j \sim \cos(j\pi/2 -\hphi(x)) e^{i\htheta(x)}$, so $C_\theta$ just takes into account the phase of the ladder operators. The extra operator $\cos(j\pi/2 -\hphi(x))$ ensures that fermions anticommute. Because we have previously identified kinks as world-lines, we expect $C_\theta(x_h-x_t,\tau_h-\tau_t)$ to be linked to the field configurations with a worm ranging from $(x_t,\tau_t)$ to $(x_h,\tau_h)$. In the following, we make this statement more explicit, and a rigorous derivation can be found in  Appendix~\ref{app:two_point}.

We first establish the discretized path integral representation of Eq.~\eqref{eq:two_point_def}. The key idea to derive it is to notice that, from the relation $[\htheta(x), \hphi(y)]=-i\frac{\pi}{2}{\rm sgn}(x-y)$, one can infer
\begin{equation}\label{eq:comm_e_itheta_phi}
    [\hphi(y),e^{\pm i\htheta(x)}]=\mp \frac{\pi}{2} {\rm sgn}(x-y) e^{\pm i\htheta(x)}.    
\end{equation}
This is interpreted as the operators $e^{\pm i \htheta(x)}$ inserting a kink of height $\pm \pi$ in the field $\phi$ at the position $x$. Consequently, the field $\phi(y,\tau^-)$ at a time right before the operator insertion at $(x,\tau)$ differs by $\mp \frac{\pi}{2} {\rm sgn}(x-y)$ from the field $\phi(y,\tau^+)$ just after.

\begin{figure}[h!]
    \centering
    \includegraphics[width=0.7\linewidth]{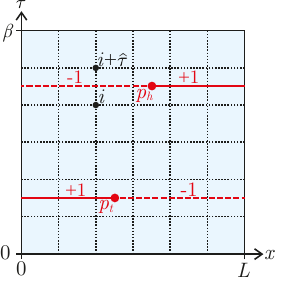}
    \caption{The insertions of the operators $e^{i\htheta(p_h)}$ and $e^{-i\htheta(p_t)}$ modify the quadratic interactions crossing the red lines as $(\phi_i - \phi_{i+\hat{\tau}})^2 \to \left(\phi_i - \phi_{i+\hat{\tau}} \pm \frac{\pi}{2}\right)^2$ where the sign is $+$ along the full lines and $-$ along the dashed ones.}
    \label{fig:C_theta_main}
\end{figure}

Putting the field theory on a lattice, its is convenient to think of the coordinates $(x_h,\tau_h)$ and $(x_t,\tau_t)$ as plaquettes $p_h$ and $p_t$ (see Fig.~\ref{fig:C_theta_main}). This is because, according to \eqref{eq:comm_e_itheta_phi}, the path integral representation of $C_\theta$ is given by the usual action $S$ from Eq.~\eqref{eq:S_discretized} except for the quadratic interactions crossing the insertion times $(p_h)_\tau$ and $(p_t)_\tau$ which are modified as
\begin{equation}
    (\phi_i - \phi_{i+\hat{\tau}})^2 \to \left(\phi_i - \phi_{i+\hat{\tau}} \pm \frac{\pi}{2}\right)^2,
\end{equation}
where the sign is explicitly given in Fig.~\ref{fig:C_theta_main}.
To find the current-fluctuation representation of this modified interaction, we first use the decomposition $\phi_i = \frac{\pi}{2}(n_i+f_i)$ which gives the interaction
\begin{equation}
    \frac{\pi^2}{4}\left(n_i - n_{i+\hat{\tau}} \pm 1 + f_i - f_{i+\hat{\tau}}\right)^2.
\end{equation}
This suggests to replace the old current definition $J^x_p=n_{i(p)+\hat{\tau}} - n_{i(p)}$ by the new one $J^x_p \equiv n_{i(p)+\hat{\tau}} - n_{i(p)} \mp 1$. The modified interaction therefore becomes
\begin{equation}
    \frac{\pi^2}{4}\left(J^x_p + f_{i(p)+\hat{\tau}} - f_{i(p)}\right)^2,
\end{equation}
which is the same as in the action \eqref{eq:current_fluc_action} already obtained for the partition function. However, from the modified definition of the current, one can show that the current field $\vec{J}$ is now divergenceless everywhere but at $p_h$ and $p_t$ where $(\vec{\nabla} \cdot \vec{J})_{p_t}=2$ and $(\vec{\nabla} \cdot \vec{J})_{p_h}=-2$. The current field therefore belongs to the ensemble $\mathcal{C}_2(p_t,p_h)$ of current configurations with a worm from $p_t$ to $p_h$, and one concludes that
\begin{align}
    C_\theta(p_h-p_t) =& \frac{\sum_{\vec{J}\in \mathcal{C}_2(p_t,p_h)} \prod_i \int_{-\frac12}^{\frac12} \d f_i \, e^{-S(\vec{J},f)}}{\sum_{\vec{J}\in \mathcal{C}_0} \prod_i \int_{-\frac12}^{\frac12} \d f_i \, e^{-S(\vec{J},f)}},
\end{align}
where the denominator is just the partition function $Z$. In terms of the average $\langle \bullet \rangle_{\rm tot.}$ with respect to the total probability distribution in the enlarged space containing the worm and the lifting variables (see also Eq.~\eqref{eq:target_distribution}), this is expressed as
\begin{align}\label{eq:two_point_bf}
    C_\theta(p) =&\frac{\langle \delta_{p_h-p_t,p}\rangle_{\rm tot.}}{\langle \delta_{p_h-p_t,0}\rangle_{\rm tot.}}.
\end{align}
This means that $C_\theta(p)$ is simply proportional to the time that the worm head spends at a distance $p$ from its tail and is normalized to give $C_\theta(0)=1$. The fact that the worm algorithm easily gives access to two-point functions is a general feature shared by many physical systems \cite{Prokofiev_1998_worm,Prokofiev_2001_worm,Alet_2003_directedWA}. A pseudocode implementation of the computation of $C_\theta(p)$ can be found in Appendix~\ref{app:pseudo_code}.

\subsection{Scaling analysis}
\label{sec:scaling_analysis}
A critical point being scale invariant, it is believed that observables should display a universal scaling behaviour in its neighbouring \cite{NdupuisCMUG2}. This so-called scaling hypothesis is not only useful for understanding the universal behaviour of observables, but is also a precise tool to analyse Monte Carlo data near criticality.

We now briefly recall the results of applying the scaling hypothesis to the generic Mott transition \cite{Fisher_1989_SI,Wallin_1994_2DSI}. The relevant physical length and time scales involved in our model are: the system size $L$ and inverse temperature $\beta$, the space and time correlation lengths $\xi$ and $\xi_\tau$, and the microscopic length of the lattice spacing. At a distance $\delta=\mu - \mu_c$ of the critical point, the correlation lengths are expected to diverge as
\begin{align}
    \xi \sim |\delta|^{-\nu}, \quad \xi_\tau \sim |\delta|^{-\nu z},
\end{align}
which defines the two critical exponents $\nu$ and $z$. The scaling hypothesis then consists in arguing that near the scale invariant critical point, the short distance details should be irrelevant and observables should only depend on $L$, $\beta$, $\xi$ and $\xi_\tau$. Therefore, an observable $\mathcal{O}$ with dimension $({\rm length})^{-[\mathcal{O}]_x} \times ({\rm time})^{-[\mathcal{O}]_\tau}$ can be expressed as
\begin{align}
    \mathcal{O}(L,\beta,\xi,\xi_\tau)=L^{-[\mathcal{O}]_x}\beta^{-[\mathcal{O}]_\tau} \mathcal{\tilde O}\left(\frac{\xi}{L},\frac{\xi_\tau}{\beta}\right),
\end{align}
which is usually written in the more convenient form
\begin{align}
    \mathcal{O}(L,\beta,\delta)=L^{-[\mathcal{O}]} \mathcal{\tilde O}\left(\delta L^{1/\nu},\frac{\beta}{L^z}\right),
\end{align}
where $[\mathcal O]=[\mathcal{O}]_x+z[\mathcal{O}]_\tau$ is the scaling dimension of $\mathcal O$, and $\mathcal{\tilde O}$ is known as a scaling function. Note that $[\mathcal{O}]$ is the full scaling dimension which may differ from the naive (or engineering) one because of anomalous dimensions coming, for instance, from the scaling field $e^{i\theta}$. We, however, do not expect any anomalous dimension for $\varphi$ and $\theta$ as they are phases, so $[\varphi]=[\theta]=0$. This assumption is, of course, verified numerically in the rest of the article. These results can be straightforwardly applied to the compressibility $\kappa$ and the superfluid stiffness $\rho_s$. From the definitions \eqref{eq:kappa_def} and \eqref{eq:rho_s_def}, their scaling dimensions are found to be
\begin{align}
    [\kappa]=&2+2\left(-\frac{z+1}{2} + [\varphi] \right)=1-z,\\
    [\rho_s]=&2z + 2\left(-\frac{z+1}{2} + [\varphi] \right)=z-1,
\end{align}
\begin{figure*}[t!]
    \centering
    \includegraphics[width=0.95\linewidth]{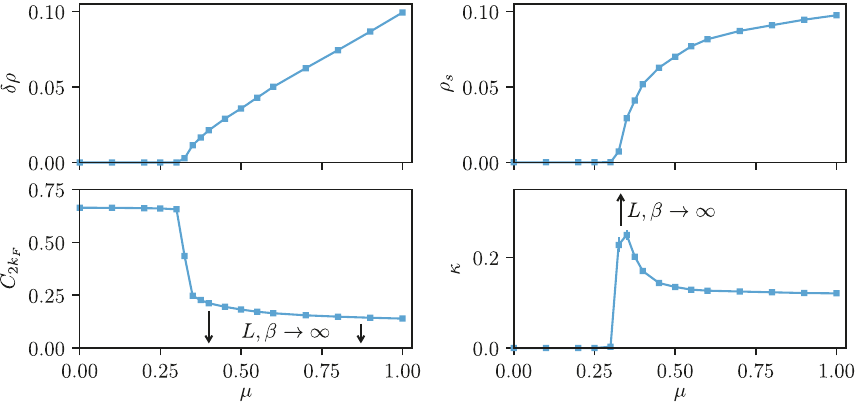}
    \caption{Top left: doping $\delta \rho$ \eqref{eq:delta_rho}. Bottom left: amplitude of the $2k_F$ density modulation $C_{2k_F}$ \eqref{eq:C_2kF}. Top right: superfluid density $\rho_s$ \eqref{eq:rho_s_def}. Bottom right: compressibility $\kappa$ \eqref{eq:kappa_def}. The data was obtained for $L=\beta = 256$. As $L,\beta$ are increased, $C_{2k_F}$ decreases in the LL, and the peak in the compressibility becomes sharper. The error bars come from averaging over multiple runs with independent initial conditions.}
    \label{fig:order_params_arrows}
\end{figure*}
from which one infers the finite-size scalings
\begin{align}\label{eq:kappa_FSS}
    \kappa(L,\beta,\delta)&= L^{z-1} \tilde \kappa \left(\delta L^{1/\nu},\frac{\beta}{L^z}\right),\\
    \label{eq:rho_s_FSS}
    \rho_s(L,\beta,\delta)&=  L^{1-z} \tilde \rho_s \left(\delta L^{1/\nu},\frac{\beta}{L^z}\right).
\end{align}
This can be exploited in Monte Carlo simulations by noting that the critical point $\mu_c$ and the exponents $\nu$ and $z$ are such that, if one scales $\beta \sim L^z$, the functions $\kappa(L,\delta) L^{1-z}$ and $\rho_s(L,\delta) L^{z-1}$ should collapse for different values of $L$ when plotted as a function of $\delta L^{1/\nu}$.

We will also study the asymptotic power-law decay of the two-point function $C_\theta(0,\tau)$. Since the field $e^{i\theta}$ is expected to be a scaling field, we must include its possible anomalous dimension $\eta$ such that $[e^{i\theta}]=(z-1+\eta)/2$. The relevant length scales now also include $\tau$, so repeating the arguments given in the previous paragraph shows that, in the $L,\beta \to \infty$ limit, one has
\begin{align}
    \label{eq:C_theta_tau}
    C_\theta(0,\tau) &= \tau^{-(z-1+\eta)/z} \tilde C_\theta^\tau(\tau |\delta|^{\nu z}).
\end{align}

\begin{figure}[h!]
    \centering
    \includegraphics[width=\linewidth]{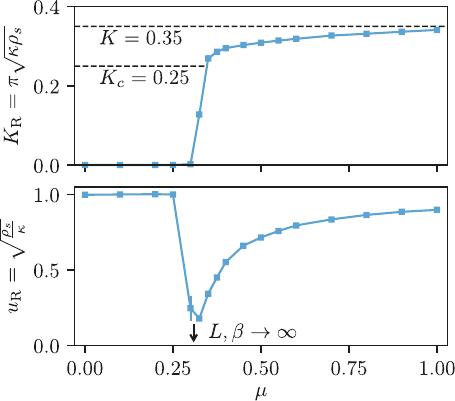}
    \caption{Luttinger parameter $K_\R$ (top) and velocity $u_\R$ (bottom) extracted from the compressibility $\kappa$ and superfluid stiffness $\rho_s$ shown in Fig.~\ref{fig:order_params_arrows}. The dip in $u_\R$ at the transition becomes deeper as the system size is increased. The data were obtained for $L=\beta = 256$. The value of $u_\R=1$ in the MI should be taken with caution as, using the MI action \eqref{eq:MI_action}, one can show that $u_\R=\sqrt{\frac{\rho_s}{\kappa}}=\frac{L}{\beta}$ and thus only reflects the relative scaling of $L$ and $\beta$. The error bars come from averaging over multiple runs with independent initial conditions.\vspace{-0.5cm}}
    \label{fig:K_u_arrows}
\end{figure}

\section{Numerical results}
\label{sec:numerical_results}
\begin{figure*}[t!]
    \centering
    \includegraphics[width=1\linewidth]{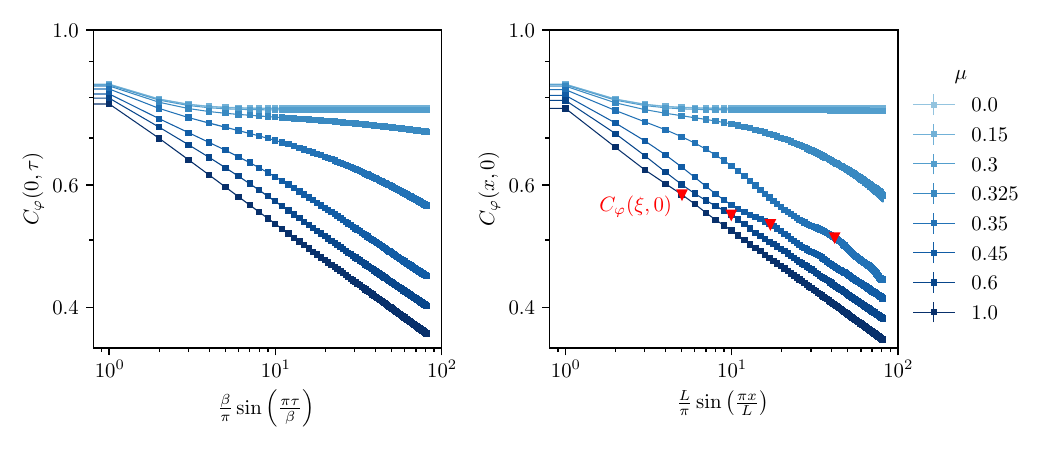}
    \caption{Two-point function $C_\varphi(x,\tau)= \langle e^{i(\varphi(x,\tau)-\varphi(0,0))}\rangle$ in space and time for several chemical potentials $\mu$ and $L=\beta=256$. We plot it against the chord functions $\frac{\beta}{\pi} \sin(\frac{\pi \tau}{\beta})$ and $\frac{L}{\pi} \sin(\frac{\pi x}{L})$ to minimize the boundary effects. The bumps in $C_\varphi(x,0)$ come from the kinks. The red points are $C_\varphi(\xi,0)$ with $\xi=1/(2\delta \rho)$ the kink-to-kink distance. The error bars come from averaging over multiple independent runs.
    }
    \label{fig:varphi_two_point}
\end{figure*}

The present section uses the SmoWo algorithm introduced previously to analyse the generic Mott transition. We first present numerical results solely for $\phi$-dependent observables, which is physically relevant when modelling commensurate-incommensurate transitions \cite{Horovitz_1983_CI_RG}, and then focus on the additional phase-phase two-point function $C_\theta$ which is specific to quantum models. In what follows we focus on $K=0.35$, $u=1$, $g=1$ and vary $\mu$ across the transition. The non-physical parameters are set to $\lambda_{\rm w}=1$ and $\lambda_{\rm r}=0.1/(\beta L)$. Typically, $10^4-10^5$ samples are generated in each run and we discard the first $10 \%$ to ensure the algorithm has thermalised.

\subsection{$\phi$-dependent observables}

\begin{figure}[h!]
    \centering
    \includegraphics[width=0.8\linewidth]{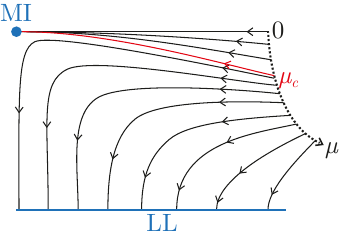}
    \caption{Schematic renormalization group flow of the tilted sine-Gordon model. We show trajectories obtained by varying the bare chemical potential $\mu$. For $\mu\le \mu_c$, they flow to the MI fixed point, while for $\mu>\mu_c$ they flow to the continuum of LL fixed points. For $\mu \to \mu_c^+$, the trajectories spend some time around the MI before heading towards the LL.}
    \label{fig:RG_flow_sG}
\end{figure}
\begin{figure*}[t!]
    \centering
    \includegraphics[width=1\linewidth]{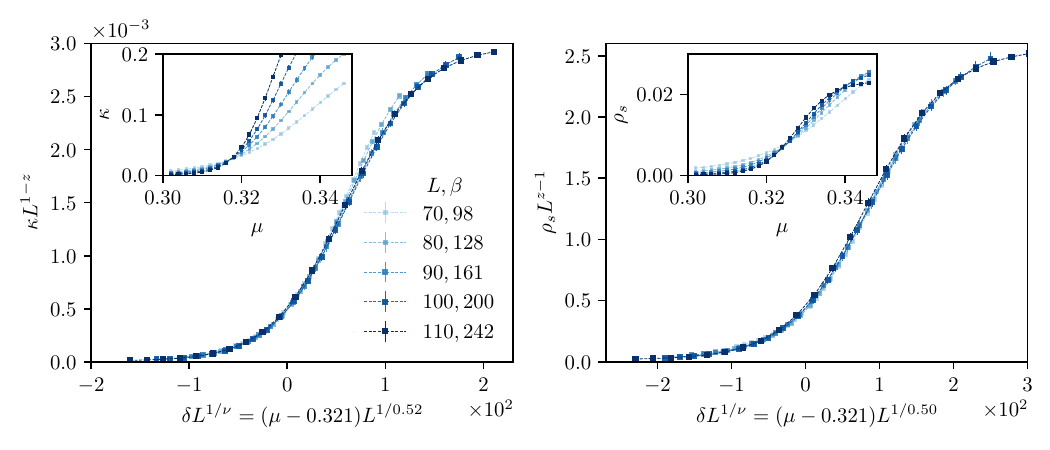}
    \caption{Finite-size scaling collapse for the compressibility $\kappa$ and superfluid stiffness $\rho_s$. The insets show the non-rescaled data while the main plots perform the rescalings suggested by Eqs.~\eqref{eq:kappa_FSS} and \eqref{eq:rho_s_FSS}. The system sizes are scaled as $\beta =L^2/50$. Superimposing the two main plots reveals that the two scaling functions for $\kappa$ and $\rho_s$ are the same up to a linear rescaling of the $x$ and $y$ axes. The error bars come from averaging over multiple independent runs.}
    \label{fig:kappa_rho_s_collapse}
\end{figure*}

The existence of two distinct phases is first checked by plotting the doping $\delta \rho$ \eqref{eq:delta_rho}, the amplitude of the $2k_F$ density modulation $C_{2 k_F}$ \eqref{eq:C_2kF}, the compressibility $\kappa$ \eqref{eq:kappa_def}, and the superfluid stiffness $\rho_s$ \eqref{eq:rho_s_def} in Fig.~\ref{fig:order_params_arrows}. We clearly identify a critical value $\mu_c$ of the chemical potential which separates two phases. For $\mu<\mu_c$, $C_{2k_F}>0$ and $\delta \rho=\kappa=\rho_s=0$, meaning the field $\phi$ gets trapped in one minimum and kinks are suppressed: this the MI. For $\mu>\mu_c$, the field $\phi$ is tilted since $\delta \rho>0$ and fluctuates a lot since $\kappa,\rho_s>0$ and $C_{2k_F}$ is small (and decreases with system size): this is the LL. We now concentrate on the LL at $\mu > \mu_c$. Its renormalised Luttinger parameter $K_\R$ and velocity $u_\R$ are extracted from the compressibility $\kappa$ and superfluid density $\rho_s$ as $K_\R= \pi \sqrt{\kappa \rho_s}$ and $u_\R=\sqrt{\rho_s/\kappa}$, and are shown in Fig.~\ref{fig:K_u_arrows}. Far from the transition point ($\mu \gg \mu_c$), the parameters $K_\R$, $u_\R$ approach their bare values $K = 0.35$ and $u=1$. Close to the transition, the Luttinger parameter $K_\R$ approaches $1/4$, as expected from analytical arguments \cite{Schulz_1980_CI,Giamarchi_1997_Mott1D}. At the same time, the velocity $u_\R$ drops to $0$ (the dip in Fig.~\ref{fig:K_u_arrows} gets sharper as the system size is increased) as $\kappa$ diverges while $\rho_s$ vanishes. This indicates a breakdown of the LL.

To understand what happens at the transition, it is useful to define a correlation length $\xi$. In the LL, it is the kink-to-kink distance $\xi=1/(2\delta \rho)$ (the factor $2$ appears because a kink is a half-particle). As for any correlation length, deep in the LL ($\mu \gg \mu_c$) $\xi \to 0$ since the kinks proliferate ($\delta \rho \to \infty$), and close to the transition $\xi \to \infty$ as kinks become scarce ($\delta \rho\to 0$). It separates small scales at which the field seems flat, and thus looks like a MI, from large ones at which the field has many kinks and appears to be a LL. This crossover is particularly visible in the two-point function $C_\varphi$ (see Fig.~\ref{fig:varphi_two_point}). Below $\xi$, $C_\varphi$ tends to be flat. Above $\xi$, $C_\varphi$ displays the typical algebraic decay of a Luttinger liquid. Notice that along the space direction, $C_\varphi(x,0)$ displays some clear bumps at $x=\xi,2\xi,3\xi,\dots$ due to the presence of kinks. In the MI, all curves are flat and collapse. The absence of any singular behaviour in $C_\varphi$ (or any of the previous observables) as $\mu \to \mu_c^-$ indicates that there is no diverging correlation length when coming from the MI and that, right at the critical point, the system is still a MI.

By only considering $\phi$-dependent observables, we have considered the physics of the sine-Gordon model rather than that of the generic Mott transition which also includes $\theta$-dependent observables. The previous analysis therefore has consequences for a renormalization group (RG) study of the sine-Gordon model. It appears that the sine-Gordon model has an RG fixed point corresponding to the MI and a continuum of LL fixed points (one for each value of $(u_\R,K_\R)$), but no intermediate critical fixed point separating both phases. Note that, strictly speaking, one has to distinguish the LL with $\delta \rho \ne 0$ (which we are studying here) from that with $\delta \rho = 0$ as the free-energy is non-analytic at $\delta \rho \to 0$ and signals the presence of a $n$th-order phase transition with $n\ge 3$ (see Ref.~\cite{Horovitz_1983_CI_RG}). Neglecting for simplicity the transitions to the LL at $\delta \rho =0$, we are lead to infer the schematic RG flow depicted in Fig.~\ref{fig:RG_flow_sG}. The trajectories obtained for $\mu \to \mu_c^+$ are first attracted to the MI, before heading towards the LL, in agreement with the fact that they represent systems whose short-distance physics looks like that of a MI. 

\begin{figure*}[t!]
    \centering
    \includegraphics[width=1\linewidth]{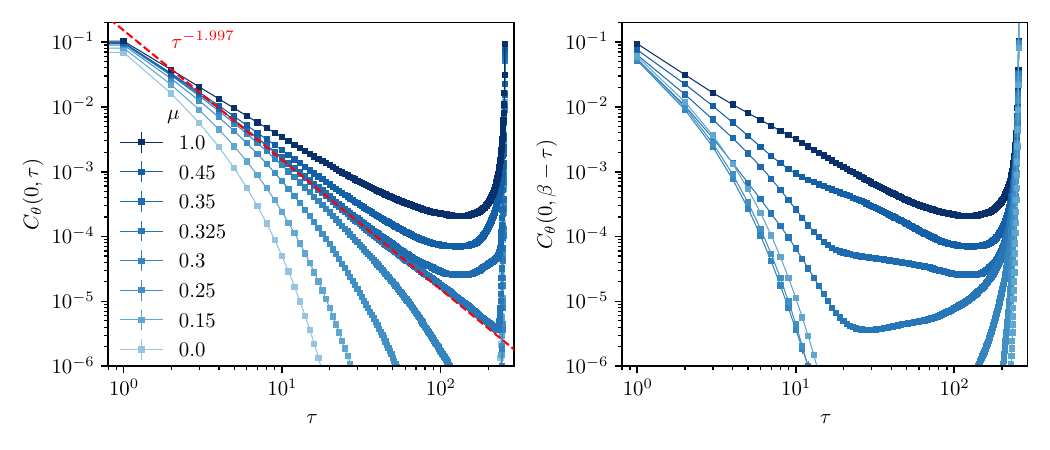}
    \caption{Two-point function $C_\theta(0,\tau)$ (left) and $C_\theta(0,\beta-\tau)$ (right) \eqref{eq:two_point_def} obtained by varying $\mu$ across the transition. The system size is $L = \beta = 256$. The error bars come from averaging over multiple independent runs. The fit in the left plot is made over $\tau \in [5,70]$.}
    \label{fig:theta_t_two_point}
\end{figure*}
Although there is no critical fixed point, it is still possible to study critical exponents since they govern the divergence of $\xi$ as $\mu \to \mu_c^+$. To characterize the critical exponents $z$, $\nu$ defined in Sec.~\ref{sec:scaling_analysis} and the critical point $\mu_c$, we use the finite-size scaling forms \eqref{eq:kappa_FSS} and \eqref{eq:rho_s_FSS}. This requires to perform simulations at different system sizes while scaling $\beta \sim L^z$, so we are forced to make a guess on $z$ before proceeding. We postulate $z=2$, a fact verified a posteriori if the scaling holds. The scaling functions to plot are thus $\kappa L^{1-z}=\kappa/L$ and $\rho_s L^{z-1}=\rho_s L$, and we scale $\beta = L^2/a$ with a large aspect ratio $a=50$ to deal with the important finite-size effects (see also the discussion in Ref.~\cite{Alet_2004_CI}). Figure~\ref{fig:kappa_rho_s_collapse} presents the finite-size scaling collapse of the observables $\kappa$ and $\rho_s$. The data were obtained by carrying simulations at $\mu=0.32$ and extrapolating them to other values of $\mu$ using reweighing techniques \cite{Ferrenberg_1988_reweighting}. The collapses in Fig.~\ref{fig:kappa_rho_s_collapse} are done using $\nu=0.50$, $\mu_c=0.321$ and $\nu=0.52$, $\mu_c=0.321$ for $\kappa$ and $\rho_s$ respectively. The error on $\nu$ is estimated as the difference between both values, while that on $\mu_c$ is determined by varying $\mu_c$ until the collapse breaks down. This leads to
\begin{equation}
    \nu = 0.51 \pm 0.01, \quad \mu_c =0.321 \pm 0.001,
\end{equation}
and the quality of the collapse is strong evidence for $z=2$. This also agrees with the relation $z \nu =1$ proven in Ref.~\cite{Fisher_1989_SI}.

\subsection{The phase-phase two-point function}

We now consider the phase-phase two-point function $C_\theta$ of 1D quantum systems. We solely concentrate on the time behaviour $C_\theta(0,\tau)$. The analysis of the full space-time dependency is more intricate and will be reported elsewhere. Since the system is not particle-hole symmetric for $\mu \ne \mu_c$, $C_\theta(0,\tau)$, which describes the phase of a particle, need not be equal to $C_\theta(0,\beta-\tau)$, which describes that of a hole. In Fig.~\ref{fig:theta_t_two_point}, left, we have plotted $C_\theta(0,\tau)$ across the transition. For $\mu>\mu_c$, we observe an algebraic decay $C_\theta(0,\tau)\sim \tau^{-1/(2K_\R)}$ typical of a Luttinger liquid. Contrary to $C_\varphi$, this algebraic decay is still present at the critical point, indicating that $e^{i\theta}$ is a scaling field whereas $e^{i\varphi}$ is not. At the critical point ($\mu=0.325 \simeq \mu_c$), fitting the algebraic time decay (for $\tau$ close to $0$, not $\beta$) gives $C_\theta(0,\tau)\sim 1/\tau^{1.997\pm 0.008}$. Assuming that $z=2$ from the previous section and using the scaling ansatz \eqref{eq:C_theta_tau} specialized to $\delta=0$, this gives an estimate of the anomalous dimension
\begin{equation}
    \eta=2.99 \pm 0.02.
\end{equation}
It is often stated that, at the critical point, the Luttinger parameter takes on the value $K_\R=1/4$ \cite{Schulz_1980_CI,Giamarchi_1997_Mott1D,Kuhner_2000_BoseHubbard_DMRG}. Since the dynamical exponent is $z=2$ at the transition, the critical point is evidently not a Luttinger liquid which has $z=1$. However, approaching it from the LL, $K_\R$ seems to converge towards $1/4$ (see Fig.~\ref{fig:K_u_arrows}) and, right at the transition, the previous observation that $C_\theta(0,\tau)\sim \tau^{-2}$ indeed corresponds to the scaling $C_\theta(0,\tau)\sim \tau^{-1/(2K_\R)}$ of a LL with $K_\R=1/4$. For $\mu<\mu_c$, $C_\theta$ gradually transitions from an algebraic decay to a faster one. The crossover time $\xi_\tau^p$ --- the superscript $p$ refers to $C_\theta(0,\tau)$ describing the propagation of a particle --- defines the previously missing correlation length $\xi=(\xi_\tau^p)^{1/z}$ which diverges as $\mu \to \mu_c^-$. 

The behaviour of $C_\theta(0,\beta-\tau)$ strongly differs from that of $C_\theta(0,\tau)$ (see Fig.~\ref{fig:theta_t_two_point}, right). In the LL ($\mu>\mu_c$), its asymptotic decay is $\sim \tau^{-1/(2K_\R)}$, just like $C_\theta(0,\tau)$, but sets in at much later times as the transition is approached. At the transition and in the MI, $C_\theta(0,\beta-\tau)\sim \exp(-\tau/\xi_\tau^h)$ where the superscript $h$ is for hole. Not only does the decay length $\xi_\tau^h$ not diverge at the transition, but it also decreases to a finite value as $\mu \to \mu_c^-$. This length can thus not be interpreted as the relevant correlation length near the transition. Its microscopic interpretation will be discussed in the next section. 

\begin{figure}[h!]
    \centering
    \includegraphics[width=1\linewidth]{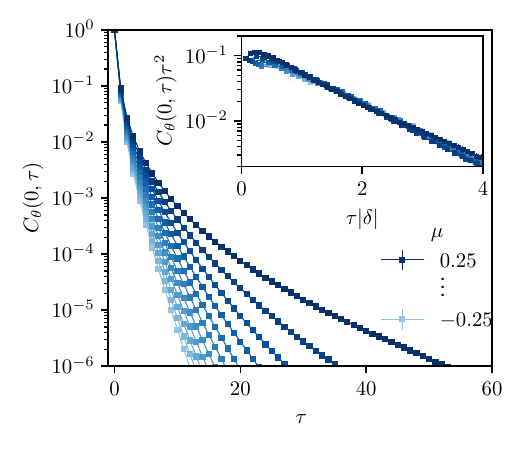}
    \caption{Two-point function $C_\theta(0,\tau)$ \eqref{eq:two_point_def} obtained by varying $\mu$ across the Mott lobe. Inset: collapse of the two-point function $C_\theta(0,\tau)$ using the scaling ansatz \eqref{eq:C_theta_tau} and the values $z=2$, $\nu=0.5$, $\eta=3$ and $\mu_c=0.321$. Note how the collapse holds even far away from the critical point. The system size is $L = \beta = 256$. The data for $\mu<0$ are extracted from that for $\mu>0$ by using $C_\theta(0,\tau;\mu)=C_\theta(0,\beta-\tau;-\mu)$ which comes from the particle-hole exchange $\theta,\phi,\mu \to -\mu,-\theta,-\phi$. The error bars come from averaging over multiple independent runs.}
    \label{fig:theta_t_Mott}
\end{figure}
\begin{figure}[h!]
    \centering
    \includegraphics[width=0.8\linewidth]{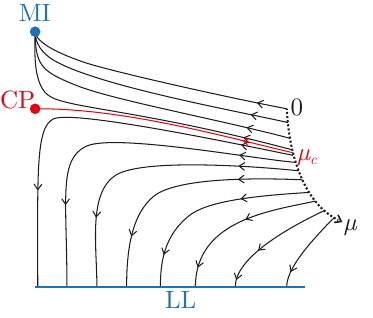}
    \caption{Schematic renormalization group flow of the enlarged model (\ref{eq:current_fluc_action},\ref{eq:MC_partition_function}) containing worms. We show trajectories obtained by varying the bare chemical potential $\mu$. For $\mu<\mu_c$, they flow to the MI fixed point, for $\mu=\mu_c$ they flow to the critical point (CP), and for $\mu>\mu_c$ they flow to the continuum of LL fixed points.}
    \label{fig:RG_flow_worm}
\end{figure}

In the MI, we have argued that the decay of $C_\theta(0,\tau)$ defines a correlation time $\xi_\tau^p$. To make this statement more precise, we plot $C_\theta(0,\tau)$ in the MI ($\mu \in [-\mu_c, \mu_c]$) in Fig.~\ref{fig:theta_t_Mott}. From the scaling ansatz \eqref{eq:C_theta_tau} and the now known values of $\nu=0.5$, $z=2$, $\eta=3$ and $\mu_c=0.321$, we collapse all curves $C_\theta(0,\tau)$ (see the inset in Fig.~\ref{fig:theta_t_Mott}) by plotting the scaling function $C_\theta(0,\tau)\tau^{(z-1+\eta)/z}=C_\theta(0,\tau)\tau^2$ as a function of $\tau /\xi_\tau^p \sim \tau |\delta|^{\nu z}=\tau|\delta|$. Since the scaling function is a line in the log-linear plot, we infer that $C_\theta(0,\tau) \sim \tau^{-(z-1+\eta)/z}\exp(-\tau/\xi_\tau^p)=\tau^{-2}\exp(-\tau/\xi_\tau^p)$ with $\xi_\tau^p \sim |\delta|^{-1}$. In terms of RG fixed points, the correlation length provided by $C_\theta$ in the MI allows distinguishing the MI where $\xi < +\infty$ from the critical fixed point (CP) where $\xi = + \infty$, leading to the schematic RG flow in Fig.~\ref{fig:RG_flow_worm}. It may seem strange that the number of fixed points should depend on the observables considered. However, by including $C_\theta$ we have actually changed our model from the sine-Gordon model (\ref{eq:partition_function},\ref{eq:phi_action}), which only contains $\phi$-dependent observables, to the enlarged model (\ref{eq:current_fluc_action},\ref{eq:MC_partition_function}) containing worms and $C_\theta$. Our results therefore show that the sine-Gordon model has two fixed points, but the enlarged model has three. As quantum models contain both $\phi$-dependent and $\theta$-dependent observables, the RG flow of the generic Mott transition should resemble that of the enlarged model and have three fixed points.

\subsection{Identifying the Hubbard bands}
\label{sec:Hubbard_bands}
\begin{figure}[h!]
    \centering
    \includegraphics[width=0.8\linewidth]{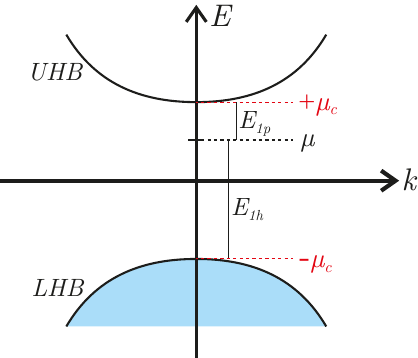}
    \caption{The MI can be described by two Hubbard bands, the upper Hubbard band (UHB) and the lower Hubbard band (LHB). At zero temperature, when the chemical potential lies in the gap, only the states in the LHB are occupied. The energy needed to create a particle is then $E_{1p}=\mu_c-\mu$, and that of a hole is $E_{1h}=\mu+\mu_c$.}
    \label{fig:Hubbard_bands}
\end{figure}

The difference between $C_\varphi(0,\tau)$, which does not change at all in the MI, and $C_\theta(0,\tau)$, which does, can be understood at the quantum level. Recall that for an operator $\mathcal{O}(\tau)$, its two-point function is given by
\begin{align}\label{eq:two_point_assymptotic}
    \langle \mathcal{O}(\tau) \mathcal{O}(0) \rangle =&\hat{\mathcal{T}} \bra{0}e^{\tau\hat{H}}\hat{\mathcal{O}}e^{-\tau\hat{H}}\hat{\mathcal{O}}\ket{0} \nonumber\\
    =& \sum_n |\bra{0}\hat{\mathcal{O}}\ket{n}|^2 e^{-|\tau| E_n}
\end{align}
where $\ket{n}$ are the energy eigenstates. This means that, at late times, the two point function decays as $\sim C + e^{-|\tau|E_{1\mathcal{O}}}$ where $E_{1\mathcal{O}}$ is the energy of the lowest-energy excitation created by $\mathcal{O}$, and $C$ is a constant that appears for observables with a non-zero average $\langle \mathcal{O}\rangle$ because of spontaneous symmetry breaking. For our purposes, we are interested in two operators. The first is the field $e^{i\varphi(x)}$ which captures the fluctuations of the density. The second is the one-body operator which, in the MI, is $c_j\sim \cos(\pi/2j-\varphi(x))e^{i\theta(x)}\simeq \cos(\pi/2j-\langle \varphi\rangle)e^{i\theta(x)}$ since the field $\varphi$ fluctuates very little. According to Eq.~\eqref{eq:two_point_assymptotic}, for $\tau \gg 1$,
\begin{align}
    C_\theta(0,\tau)\sim e^{-\tau E_{1p}},
\end{align}
where $E_{1p}$ is the smallest energy needed to create a particle. Since it sets the asymptotic exponential decay of $C_\theta(0,\tau)$, we identify $E_{1p} = 1/\xi_\tau^p$. Similarly, for the propagation of a hole, $C_\theta(0,\beta-\tau) \sim e^{-\tau E_{1h}}$ with
$E_{1h}\sim 1/\xi_\tau^h$ the smallest energy needed to create a hole. We also have
\begin{align}
    C_\varphi \sim C + e^{-\tau E_{1p1h}},
\end{align}
where $E_{1p1h}$ is the smallest energy needed to create a density excitation, i.e., a particle and a hole, and $C$ reflects the spontaneous breaking of the symmetry $\varphi \to \varphi + \pi/2$, which microscopically consists in reversing the "particle-hole-particle-hole" order of the MI to "hole-particle-hole-particle". The observation that $C_\varphi(0,\tau)$ does not depend on $\mu$ while $C_\theta(0,\tau)$ does, implies that $E_{1p1h}$ does not vary with $\mu$ whereas $E_{1p}$ does. This agrees with the picture of the MI having an upper Hubbard band with energies $E>\mu_c$ and a lower Hubbard band with $E< - \mu_c$ separated by a gap $2 \mu_c$ (see Fig.~\ref{fig:Hubbard_bands}) \cite{Giamarchi_1997_Mott1D}.
When the chemical potential $\mu$ is in between the two bands, the energy needed to create a particle is $E_{1p}=\mu_c - \mu=|\delta|$ which agrees with $\xi_\tau^p \sim |\delta|^{-1}$ and explains why the critical scaling \eqref{eq:C_theta_tau} holds throughout the MI and not just close to the critical region. The energy of a hole being $E_{1h}=\mu + \mu_c$, it increases to the finite value $2\mu_c$ at the transition. As soon as $\mu>\mu_c$, particles start to populate the upper Hubbard band and $E_{1h}=0$. This discontinuity in $E_{1h}$ is responsible for the large crossover in the LL between the early exponential decay and the late algebraic decay (see Fig.~\ref{fig:theta_t_two_point}, right). Finally, the energy of a particle and a hole is $E_{1p1h}=E_{1p}+E_{1h}=2 \mu_c$, which, as expected, is $\mu$-independent. The MI can also be exited through $-\mu_c$ instead of $+\mu_c$. The roles of $\xi_\tau^h$ and $\xi_\tau^p$ would then be reversed, with $\xi_\tau^h$ diverging at the transition.

\section{Conclusion}
\label{sec:conclusion}
In this work, we have proposed a Monte Carlo algorithm to study the generic Mott transition in one-dimensional quantum systems. More specifically, we considered the bosonized formulation of such models which, in the path-integral formalism, is given by the sine-Gordon model with a tilt \eqref{eq:phi_action}. In the grand-canonical ensemble, the boundary conditions of this model are only periodic modulo $\pi$. This splits the configuration space into distinct topological sectors, rendering large-scale simulations totally out of reach for standard Metropolis-like algorithms. The key idea to build an efficient algorithm is thus to enlarge the model and work in a configuration space where one can smoothly transition between different boundary conditions. This idea has already been successfully used to study Mott transitions in $(2+1)$D \cite{Alet_2003_directedWA,Alet_2003_geometricWA,Alet_2004_CI,Prokofiev_2004_SI_dis} by introducing  worm updates \cite{Prokofiev_1998_worm,Prokofiev_2001_worm} that gradually change the boundary conditions. However, these studies were made on very coarse-grained models whose $(1+1)$D counterpart would be obtained by neglecting the fine fluctuations in the sine-Gordon model. To efficiently deal with those fluctuations, we introduced a smooth worm algorithm (SmoWo) which combines the previous worm updates with event-chain Monte Carlo updates \cite{Michel2014GenECMC,krauth2021ECMC}. Both types of updates are tightly intertwined so as to smooth out each worm move and push even further the idea of generating a smooth path between configurations with different boundary conditions. We performed a detailed performance analysis and showed that the SmoWo algorithm has an integrated autocorrelation time which scales as $O(L^{4.8})$ for a system of size $\sim L \times L^2$. More concretely, simulating a $256\times 256$ system with the SmoWo algorithm takes roughly the same CPU time as simulating a $15\times15$ system with a naive Metropolis--Hastings algorithm.

Although models describing generic Mott transitions are known to have very large finite-size effects \cite{Alet_2004_CI,Prokofiev_2004_SI_dis}, the SmoWo algorithm proved powerful enough to clearly identify the two phases: the Luttinger liquid with quasi-long range order, and the Mott insulator with its characteristic Hubbard bands. At the critical point, we also retrieved the exponents $z=2.00 \pm 0.01$, $\nu=0.51\pm 0.01$ and the anomalous dimension for the particle-particle two-point function $\eta=2.99 \pm 0.02$, all of which compare favourably with analytical results derived from the Bethe ansatz or refermionization techniques \cite{Yamamoto_1983_sG,Haldane_1983,Schulz_1980_CI,Papa_2001_sG}.

While the present work focused on simulating the sine-Gordon model which is by now relatively well understood, the SmoWo algorithm developed here is relevant for the study of any bosonized system in the grand-canonical ensemble --- that is, provided the Euclidean action contains no imaginary part, which is the case if one can integrate out one of the bosonic fields $\phi$ or $\theta$ --- and thus complements the algorithm proposed in Ref.~\cite{bouvdup_2025_bosonized1d} for bosonized systems in the canonical ensemble. For instance, the SmoWo algorithm can naturally be extended to study more complicated models including long-range interactions \cite{Orignac_2025}, dissipative effects \cite{Malatsetxebarria_2013}, or multiple fields \cite{Paris_2025}.

\begin{center}
{\bf ACKNOWLEDGEMENTS}
\end{center}
We thank Edmond Orignac for insightful discussions.
O.B.-D. acknowledges the support of the French ANR under the grant ANR-22-CMAS-0001 (\emph{QuanTEdu-France} project). O. B.-D. also thanks Bruzon, where a significant part of this work was done, for its hospitality.
This work was supported by the French government through the France 2030 program (PhOM – Graduate School of Physics), under reference ANR-11-IDEX-0003 (Project Mascotte, L. Foini).

\begin{center}
{\bf DATA AVAILABILITY}
\end{center}
The data and code that support the findings of this article
are openly available \footnote{\url{https://github.com/OscarBouverot-Dupuis/algo-phys/tree/master/embedded-worm-Mott-bosonized}}.

\appendix
\begin{widetext}

\section{Bosonization subtleties}
\label{app:Bosonization_subtleties}
In this appendix, we explain in greater detail the bosonization approach used in this work. We first recall our bosonization conventions as there are many competing conventions used in the literature, and we pay particular attention to the treatment of the so-called 0-modes. Next, we explain how to obtain the path integral formulation of the bosonized Hamiltonian \eqref{eq:H_bosonized} and the phase-phase two-point function \eqref{eq:two_point_def}, and finally comment on the bosonized definitions of some observables.

\subsection{Bosonization conventions}
Our bosonization conventions follow that of constructive abelian bosonization as presented in \cite{Haldane_1981,von_Delft_bosonization} (see also Appendix B of \cite{Chua_2020}). For a periodic system of length $L$ and Fermi momentum $k_F$, this approach requires first splitting the fermionic field $\hpsi(x=aj)=c_j/\sqrt{a}$ into $\sum_{r= \pm 1} \hpsi_r$ to distinguish left-movers $\hpsi_{r=-1}$ with momenta close to $-k_F$ from right-movers $\hpsi_{r=+1}$ with momenta close to $+k_F$. The bosonization identity then states that these fermionic degrees of freedom can be expressed using bosonic fields $\hphi_r$ as
\begin{align}\label{eq:bosonization_identity}
    \hpsi_r(x)=&\frac{\hat{\eta}_r}{\sqrt{L}} :e^{i r k_F x - i\hphi_r(x)}:,
\end{align}
where the bosonic fields $\hphi_r$ are defined through their mode decomposition
\begin{align}
    \hphi_r(x)=\hphi_r^0 - r2\pi\hat{N}_r \frac{x}{L}  +i\sum_{r q>0}\sqrt{\frac{2 \pi}{L|q|}} \left(e^{iq x}\hb_r(q) -e^{-iq x}\hb_r^\dagger(q)\right),
\end{align}
with $q \in \frac{2\pi}{L}\mathbb{Z}$, and $:\bullet:$ denotes bosonic normal-ordering (all operators $\hphi_r^0,\hb^\dagger(q)$ are to be put to the left of all $\hat{N}_r,\hb_r(q)$). The operators $\hphi_r^0$ are the 0-momentum modes ($q=0$) and are only defined in terms of the original fermions through $e^{i\hphi_r^0}$. Thus, one has to identify $\hphi_r^0 \sim \hphi_r^0 + 2\pi$, which means that the fields $\hphi_r(x)$ are compact bosons with compactification radii $2\pi$. The integer-valued operators $\hat{N}_r$ count the number of $r$-movers. The $\hb_r(q),\hb^\dagger_r(q)$ correspond to the bosonic particle-hole excitations of the system. Finally, the $\hat{\eta}_r$ in Eq.~\eqref{eq:bosonization_identity} are Majorana fermions which ensure the anti-commuting nature of the fermion. The only non-zero commutators amongst these operators are
\begin{align}
    [\hb_r(q),\hb^\dagger_{r'}(q')]=\delta_{rr'}\delta_{qq'},\quad [\hphi_r^0,\hat{N}_{r'}]=i\delta_{r,r'}, \quad \{\hat{\eta}_r,\hat{\eta}_{r'} \}=2\delta_{r,r'}.
\end{align}
The phase fields introduced in the main text are given by ($R$ is for $r=1$ and $L$ for $r=-1$)
\begin{align}
    \hphi(x)=\frac{\hphi_R(x)-\hphi_L(x)}{2}, \qquad \htheta(x)=-\frac{\hphi_R(x)+\hphi_L(x)}{2}.
\end{align}
They can be decomposed as
\begin{align}\label{eq:phi_theta_decomposition}
    \hphi(x)=&\hphi_0 - \frac{\pi \hat{N} x}{L} + \tilde \phi(x), \quad \htheta(x)=\htheta_0 + \frac{\pi \hat{J} x}{L} + \tilde \theta(x) ,
\end{align}
where the topological operators are
\begin{alignat}{2}
    \hphi_0=&\frac{\hphi_R^0-\hphi_L^0}{2}, \qquad &&\htheta_0=-\frac{\hphi_R^0 + \hphi_L^0}{2},\\
    \hat{N}=&\hat{N}_R+\hat{N}_L, \qquad &&\hat{J}=\hat{N}_R - \hat{N}_L,
\end{alignat}
and the $L$-periodic operators are
\begin{align}
    \tilde \phi(x)=&i\sum_{q\ne 0}\sqrt{\frac{\pi}{2 L |q|}}{{\rm sgn}(q)} \left(e^{iq x}\hb_{{\rm sgn}(q)}(q) -e^{-iq x}\hb_{{\rm sgn}(q)}^\dagger(q)\right),\\
    \tilde \theta(x)=&-i\sum_{q \ne 0}\sqrt{\frac{\pi}{2L|q|}} \left(e^{iq x}\hb_{{\rm sgn}(q)}(q) -e^{-iq x}\hb_{{\rm sgn}(q)}^\dagger(q)\right).
\end{align}
The only non-vanishing commutation relations are
\begin{equation}\label{eq:commutation_relations}
    [\hat{N},\htheta_0]=i, \quad [\hphi_0,\hat{J}]=i, \quad  [\tilde \theta(x),\tilde \phi(y)]=-i\pi\left(\left\lfloor\frac{x-y}{L} \right\rfloor + \frac12 \right).
\end{equation}
The total particle number operator $\hat{N}$ and current operator $\hat{J}$ are integer-quantized and subject to the parity condition $\hat{N}+\hat{J} \equiv 0 \pmod{2}$. The $0$-modes $\htheta_0$ and $\hphi_0$ are compactified with radii $\pi$ but are not independent: one has to simultaneously identify $\hphi_0,\htheta_0 \sim \hphi_0+\pi,\htheta_0+\pi$ or $\hphi_0,\htheta_0 \sim \hphi_0+\pi,\htheta_0-\pi$ \footnote{Note that the compactification conditions are also seen in the harmonic fluid approach or phenomenological bosonization of bosons \cite{Haldane_1981,Cazalilla_2004_harmonic_fluid,1D_bosons_review,Giamarchi}.}. The last commutation relation implies that $[\tilde \theta(x),\frac{1}{\pi}\nabla \tilde \phi(y)]=i\sum_{n\in \mathbb{Z}}\delta(x-y-nL)$ so $\tilde \theta(x)$ and $\frac{1}{\pi}\nabla \tilde \phi(y)$ form a pair of canonically conjugate variables. Note that in the thermodynamic limit $L\to \infty$, one recovers the commutator given in the main text
\begin{equation}
    [\htheta(x), \hphi(y)]=i\pi\left( \frac{x-y}{L}-\left\lfloor\frac{x-y}{L} \right\rfloor - \frac12 \right) \overset{L\to \infty}{\longrightarrow} -i\frac{\pi}{2}{\rm sgn}(x-y).
\end{equation}
Neglecting subleading terms $\propto 1/L$ is common practice and is, of course, valid only if one is interested in the thermodynamic limit. Since this is the case in this work, we will systematically eliminate all such terms as they lead to lengthy equations. As a consequence, the simulations done at finite $L$ do not strictly coincide with ``true" finite systems of size $L$, but they nevertheless converge to the correct thermodynamic behaviour as $L$ is increased.

\subsection{Path integral formulation of the partition function}
Using the previous bosonization dictionary, the fermion \eqref{eq:H_fermion} and spin \eqref{eq:H_spin} Hamiltonians are mapped onto the sine-Gordon Hamiltonian \eqref{eq:H_bosonized} which we recall here
\begin{align}
    \hat{H}=\int_0^L \d x \frac{u}{2 \pi}\left[ K(\nabla\htheta)^2 + \frac{1}{K}(\nabla\hphi)^2\right] - g \cos(4\hphi) +\frac{\mu}{\pi}\nabla \hphi,
\end{align}
Details of the derivation of this standard result can be found in Refs.~\cite{Giamarchi,Gogolin_2004,NdupuisCMUG2}. We stress that this result only holds if one neglects contributions $\propto 1/L$ in the spirit of the present work. Using the notations introduced in Eq.~\eqref{eq:phi_theta_decomposition} and using the fact that $\tilde \phi(x)$ and $\tilde \theta(x)$ are periodic operators, the sine-Gordon Hamiltonian can be rewritten as
\begin{align}
    \hat{H}=\frac{\pi u}{2 L}\left( K \hat{J}^2 + \frac{1}{K}\hat{N}^2\right)- \mu \hat{N}+\int_0^L \d x \frac{u}{2 \pi}\left( K(\nabla \tilde \theta)^2 + \frac{1}{K}(\nabla \tilde \phi)^2\right) - g \cos\left(4\hphi_0 - \frac{4 \pi \hat{N} x}{L} + 4\tilde \phi\right).
\end{align}
We now wish to represent the canonical equilibrium partition function $Z={\rm Tr}\, e^{-\beta \hat{H}}$ using a path-integral. Since the standard procedure of Trotterizing the partition function as $Z=\lim_{M \to \infty} {\rm Tr} \prod_{i=1}^M e^{-\beta/M \hat{H}}$ always leads in the end to defining the Euclidean action as $S=\int \d \tau L$ with $L$ the classical Lagrangian in imaginary time, we decide to directly write down $L$. From Eq.~\eqref{eq:commutation_relations}, it appears that the conjugate momenta to the variables $\hat{X}=\hat{N},\hphi_0,\tilde \theta(x)$ are $\hat{P}=\htheta_0,\hat{J},\frac{1}{\pi}\nabla \tilde \phi$. Using the notation
\begin{equation}
    \partial_t \hat{X} = \frac{\delta \hat{H}}{\delta \hat{P}},
\end{equation}
the real-time Lagrangian is defined as
\begin{align}
    L=&\theta_0 \partial_t N + J \partial_t \phi_0 + \int \d x  \frac{1}{\pi} \nabla \tilde \phi \partial_t \tilde \theta - H.
\end{align}
Going to imaginary time $\tau=it$, the Euclidean action is thus given by
\begin{align}\label{eq:phi_theta_action}
    S=&\int_0^\beta \d \tau \Bigg[\frac{\pi u}{2 L}\left( K J^2 + \frac{1}{K} N^2\right)- \mu N - i\theta_0 \partial_\tau N - i J \partial_\tau \phi_0 \nonumber\\
    &+\int_0^L \d x \frac{u}{2 \pi}\left( K(\nabla \tilde \theta)^2 + \frac{1}{K}(\nabla \tilde \phi)^2\right) - g \cos\left(4\phi_0 - \frac{4 \pi N x}{L} + 4\tilde \phi\right) - i \frac{1}{\pi} \partial_x \tilde \phi \partial_\tau \tilde \theta\Bigg],
\end{align}
and the grand-canonical equilibrium partition function $Z={\rm Tr} \, e^{-\beta \hat{H}}$ is
\begin{align}
    Z=\int \mathcal{D}\tilde \phi \mathcal{D}\tilde \theta \mathcal{D} N \mathcal{D} J \mathcal{D}\phi_0 \mathcal{D}\theta_0 \, e^{-S},
\end{align}
where, following the considerations of the previous subsection, we integrate over
\begin{itemize}
    \item all space- and time-periodic functions $\tilde \phi(x,\tau)$ and $\tilde \theta(x,\tau)$,
    \item all time-periodic integer functions $N(\tau)$ and $J(\tau)$ such that $N(\tau) + J(\tau) \equiv 0 \pmod{2}$,
    \item all functions $\theta_0(\tau)$ and $\phi_0(\tau)$ such that $\theta_0(0)=N_\tau' \pi + \theta_0(\beta)$ and $\phi_0(0)=N_\tau \pi + \phi_0(\beta)$ with $N_\tau + N_\tau' \equiv 0 \pmod{2}$.
\end{itemize}
The path-integrals over $\tilde \theta$ and $\theta_0$ can be done analytically. The first integration creates a term $\sim (\partial_\tau \tilde \phi)^2$ and the second yields the condition $\partial_\tau N(\tau)=0 \Rightarrow N(\tau)=N_x$ in agreement with the fact that at the Hamiltonian level $[\hat{N},\hat{H}]=0$. Next, for $L$ large, $j=J/L$ loses its discrete nature and can be replaced by a continuous real variable up to $\propto 1/L$ corrections that we neglect. Integrating out this variable creates a term $\sim (\partial_\tau \phi_0)^2$. Putting everything together, the action reduces to
\begin{align}
    S=&\int \d x \,\d \tau \Bigg[\frac{1}{2 \pi K}\left( u(\partial_x \tilde \phi)^2 + u \left(\frac{\pi N_x}{L}\right)^2 + \frac1u (\partial_\tau \tilde \phi)^2 + \frac1u (\partial_\tau \tilde \phi_0)^2\right) - g \cos\left(4\phi_0 - \frac{4 \pi N_x x}{L} + 4\tilde \phi\right) - \mu \frac{N_x}{L}\Bigg]\nonumber\\
    =&\int \d x \,\d \tau  \Bigg[\frac{1}{2 \pi K}\left( u(\partial_x \phi)^2 + \frac1u (\partial_\tau \phi)^2\right) - g \cos\left(4\phi\right)+\frac{\mu}{\pi}\partial_x \phi\Bigg],
\end{align}
where $\phi(x,\tau)=\phi_0(\tau)-\pi N_x \frac{x}{L} + \tilde \phi(x,\tau)$. The partition function can now be explicitly written as
\begin{align}
    Z = \sum_{N_x,N_\tau=-\infty}^{+\infty} \underset{\substack{\phi(0,\tau)=\phi(L,\tau)+N_x \pi\\ \phi(x,0)=\phi(x,\beta)+N_\tau \pi}}{\int \mathcal{D}\phi} e^{-S[\phi]},
\end{align}
which is the result given in Sec.~\ref{sec:model}.

\subsection{Path integral formulation of the phase-phase two-point function}
\label{app:two_point}
This appendix proves that the discretized path integral representation of the phase-phase two-point function $C_\theta(x,\tau)=\big\langle e^{i(\theta(x,\tau)-\theta(0,0))}\big\rangle$ is Eq.~\eqref{eq:S_mod}. To this end, we first prove the following \emph{gluing identity},
\begin{equation}\label{eq:gluing_identity}
    \bra{\phi_1}e^{\pm i\htheta(x)}\ket{\phi_2} = \delta\left(\phi_1-\phi_2 \pm \frac{\pi}{2} {\rm sgn}(x-\bullet )\right),
\end{equation}
where the Dirac delta requires that, for all $y$, $\phi_1(y) -\phi_2(y) \pm \frac{\pi}{2} {\rm sgn}(x-y)=0$.

{\it Proof of the gluing identity.}
Starting from the commutation relation $[\htheta(x), \hphi(y)]=-i\frac{\pi}{2}{\rm sgn}(x-y)$, one infers the relation $[\hphi(y),e^{\pm i\htheta(x)}]=\mp \frac{\pi}{2} {\rm sgn}(x-y) e^{\pm i\htheta(x)}$. With the help of the eigenstates and eigenvalues $\ket{\phi},\phi(x)$ of the operator $\hphi(x)$ defined through $\hphi(x)\ket{\phi}=\phi(x)\ket{\phi}$, the previous operator identity yields
\begin{align}\label{eq:matrix_elem1}
    &\bra{\phi_1}\hphi(y) e^{\pm i\htheta(x)}\ket{\phi_2} = \left(\phi_2(y) \mp \frac{\pi}{2} {\rm sgn}(x-y)\right)\bra{\phi_1}e^{\pm i\htheta(x)}\ket{\phi_2}.  
\end{align}
Acting directly with $\hphi(y)$ on $\bra{\phi_1}$, the matrix element $\bra{\phi_1}\hphi(y) e^{\pm i\htheta(x)}\ket{\phi}$ can also be evaluated to
\begin{equation}\label{eq:matrix_elem2}
    \bra{\phi_1}\hphi(y) e^{\pm i\htheta(x)}\ket{\phi_2} = \phi_1(y) \bra{\phi_1} e^{\pm i\htheta(x)}\ket{\phi_2}.
\end{equation}
Putting together Eqs.~\eqref{eq:matrix_elem1} and \eqref{eq:matrix_elem2} yields
\begin{align}
    \left( \phi_1(y)  - \phi_2(y) \pm \frac{\pi}{2} {\rm sgn}(x-y)\right)\bra{\phi_1}e^{\pm i\htheta(x)}\ket{\phi_2}=0.
\end{align}
Since this identity is valid for all $y$, $\bra{\phi_1}e^{\pm i\htheta(x)}\ket{\phi_2}$ can only be non-zero when $\phi_1(y)  - \phi_2(y) \pm \frac{\pi}{2} {\rm sgn}(x-y)=0$ for all $y$. This means that
\begin{equation}
    \bra{\phi_1}e^{\pm i\htheta(x)}\ket{\phi_2} = \mathcal{N}^{(\star)} \, \delta\left(\phi_1 - \phi_2 \pm \frac{\pi}{2} {\rm sgn}(x-\bullet )\right),
\end{equation}
with a normalization factor $\mathcal{N}$. One then notices that
\begin{align}
    \delta(\phi_1-\phi_3)=\braket{\phi_1 | \phi_3}=&\int \d \phi_2 \bra{\phi_1}  e^{ i\htheta(x)}\ket{\phi_2} \bra{\phi_2}  e^{-i\htheta(x)}\ket{\phi_3}\nonumber\\
    =& \int \d \phi_2 \mathcal{N} \, \delta\left(\phi_1-\phi_2 + \frac{\pi}{2} {\rm sgn}(x-\bullet )\right) \mathcal{N}^{\star} \, \delta\left(\phi_2 - \phi_3 - \frac{\pi}{2} {\rm sgn}(x-\bullet )\right)\nonumber\\
    =& |\mathcal{N}|^2 \delta\left(\phi_1 -\phi_3 \right),
\end{align}
so $\mathcal{N}=e^{i\alpha}$. Since the matrix elements $\bra{\phi_1}e^{\pm i\htheta(x)}\ket{\phi_2}$ appear as conjugate pairs in the following, only $|\mathcal{N}|^2$ appears and one can safely ignore $\alpha$. This leads to the gluing condition Eq.~\eqref{eq:gluing_identity} stated above.

The path integral representation of $C_\theta$ defined in Eq.~\eqref{eq:two_point_def} is found by first inserting resolutions of the identity as
\begin{align}
    C_\theta(x_h - x_t, \tau_h - \tau_t) = \frac{1}{Z} \int \d \phi_1 \d \phi_2 \d \phi_3 \d \phi_4\,& \bra{\phi_1}  e^{-(\beta+\tau_t - \tau_h) \hat{H}} \ket{\phi_2}\bra{\phi_2} e^{i\htheta(x_h)} \ket{\phi_3}\nonumber\\
    &\times \bra{\phi_3} e^{-(\tau_h-\tau_t)\hat{H}} \ket{\phi_4}\bra{\phi_4} e^{-i\htheta(x_t)}\ket{\phi_1},
\end{align}
where, for the sake of simplicity, we consider $\tau_h > \tau_t$ to get rid of time-ordering. We also use the eigenstates and eigenvalues $\ket{\phi},\phi(x)$ of the operator $\hphi(x)$ defined through $\hphi(x)\ket{\phi}=\phi(x)\ket{\phi}$. The matrix elements of the operators $e^{\pm i \htheta}$ "glue" the boundary conditions of the evolution operators according to the gluing identity\eqref{eq:gluing_identity}. This leads to
\begin{align}\label{eq:C_theta_matrix_elem}
    C_\theta(x_h - x_t, \tau_h - \tau_t) = \frac{1}{Z} \int \d \phi_2 \d \phi_4 \,& \bra{\phi_4 - \frac{\pi}{2}{\rm sgn}(x_t-\bullet)}  e^{-(\beta+\tau_t - \tau_h) \hat{H}} \ket{\phi_2}\nonumber\\
    &\times \bra{\phi_2 + \frac{\pi}{2}{\rm sgn}(x_h-\bullet)} e^{-(\tau_h-\tau_t)\hat{H}} \ket{\phi_4}.
\end{align}
The path integral representations of the evolution operator matrix elements are found similarly to that of the partition function \eqref{eq:partition_function} and are
\begin{align}\label{eq:path_int_evolution_op}
    \bra{\phi_1} e^{-\tau \hat{H}} \ket{\phi_2} = \sum_{N_x=-\infty}^{+\infty}\underset{\substack{\phi(0,\tau)=\phi(L,\tau)+N_x \pi\\ \phi(x,0)=\phi_1(x)\\ \phi(x,\tau)=\phi_2(x)}}{\int \mathcal{D}\phi} e^{-S[\phi]},
\end{align}
where $S$ is the action \eqref{eq:phi_action} and $\phi$ is defined over a space time of size $L \times \tau$. Inserting Eq.~\eqref{eq:path_int_evolution_op} into \eqref{eq:C_theta_matrix_elem} yields the total path integral representation
\begin{align}\label{eq:path_int_C_theta}
    &C_\theta(x_h - x_t, \tau_h - \tau_t) = \frac{1}{Z} \sum_{N_x,N_\tau = -\infty}^{+\infty} \int \mathcal{D}\phi \, e^{-S[\phi]},
\end{align}
where the sum over $N_\tau$ is added by remembering that the boson $\phi$ is compact, i.e., $\phi$ and $\phi+\pi$ carry the same physical meaning. The path integral is over all fields $\phi$ with the boundary conditions
\begin{equation}\label{eq:boundary_phi_disc1}
    \phi(x,0)=\phi(x,\beta)+N_\tau \pi,\qquad \phi(0,\tau)=\phi(L,\tau)+ \begin{cases}
        (N_x+1)\pi \text{ if } \tau \in [\tau_t,\tau_h],\\
        N_x\pi \text{ otherwise,}
    \end{cases}\hspace{-0.4cm}
\end{equation}
and the discontinuities
\begin{align}\label{eq:disc_phi_disc1}
    \phi(x,\tau_t^+)=\phi(x,\tau_t^-)+\frac{\pi}{2} {\rm sgn}(x_t-x),\qquad \phi(x,\tau_h^+)=\phi(x,\tau_h^-) - \frac{\pi}{2} {\rm sgn}(x_h-x).
\end{align}
Conditions \eqref{eq:boundary_phi_disc1} and \eqref{eq:disc_phi_disc1} are summarized in Fig.~\ref{fig:C_theta_discr}, left.
\begin{figure}[h!]
    \centering
    \includegraphics[width=0.7\linewidth]{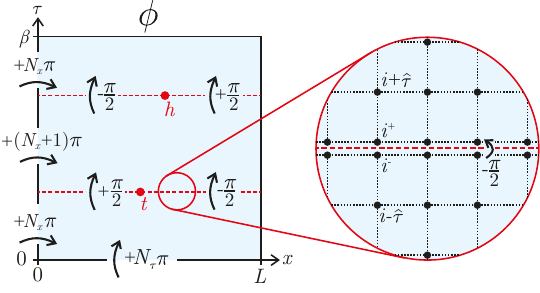}
    \caption{(Left) Boundary conditions and discontinuities \eqref{eq:boundary_phi_disc1} and \ref{eq:disc_phi_disc1} appearing in the field $\phi$. Note the similarity with the field representation of the worm configuration in Fig.~\ref{fig:loop_worm}, top. (Right) In the discretized field theory, the fields above and below the discontinuities are identified up to $\pm \pi/2$. In the picture above, one has $\phi_{i^+}=\phi_{i^-}-\pi/2$.}
    \label{fig:C_theta_discr}
\end{figure}

The path integral \eqref{eq:path_int_C_theta} consists of two fields --- one in between the discontinuities and one outside --- each weighted by the action $S$ and glued at $\tau_t$ and $\tau_h$. Both these fields can be discretized to become weighted by the discretized action \eqref{eq:S_discretized}. At the discontinuities between both fields, taking for example the situation depicted in Fig.~\ref{fig:C_theta_discr} right, one must identify the sites above and below the discontinuity as $\phi_{i^+}=\phi_{i^-}-\pi/2$. Setting $\phi_i=\phi_{i^+}$, this site interacts with $\phi_{i-\hat{\tau}}$ through
\begin{equation}\label{eq:LL_disc}
    \frac{1}{2\pi K}(\phi_{i^-}-\phi_{i-\hat{\tau}})^2=\frac{1}{2\pi K}\left(\phi_i- \phi_{i-\hat{\tau}}+\frac{\pi}{2}\right)^2,
\end{equation}
and interacts normally, i.e. with the action \eqref{eq:S_discretized}, with all other sites. Since the discontinuities affect the interactions between nearest neighbours, it is convenient to think of them as living on the edges of the dual lattice, so the points $x_h$, $x_t$ become plaquettes $p_h$, $p_t$. Extending this analysis to all sites near the discontinuities shows that the fusion of the two discretized fields follows the modified action
\begin{align}\label{eq:S_mod}
    S_{\rm mod.}(\phi)=&\sum_i \frac{1}{2 \pi K}\Big[(\phi_{i+\hat{x}} -\phi_i )^2 + \left(\phi_{i+\hat{\tau}} - \phi_i + \frac{\pi}{2} \mathbbm{1}(i,i+\hat{\tau})\right)^2\Big] - g\cos(4\phi_i)+\frac{\mu}{\pi}\left(\phi_{i+\hat{x}}-\phi_i\right),
\end{align}
where $\mathbbm{1}(i,i+\hat{\tau})$ is given by
\begin{equation}
    \mathbbm{1}(i,i+\hat{\tau})=
    \begin{cases}
        -{\rm sgn}(i_x - (p_h)_x) &\text{ if }i_\tau+\frac12=(p_h)_\tau,\\
    + {\rm sgn}(i_x - (p_t)_x) &\text{ if } i_\tau+\frac12=(p_t)_\tau,\\
    0 &\text{ otherwise}.
    \end{cases}
\end{equation}
This result was derived using $\tau_h > \tau_t$ but doing the same computation for $\tau_h < \tau_t$ recovers the same result if one replaces $[\tau_t ,\tau_h]$ in Eq.~\eqref{eq:boundary_phi_disc1} by $[\tau_t, \tau_h+\beta]$ which wraps around the periodic imaginary-time direction.

\subsection{Compressibility and superfluid density}
\label{app:kappa_rho_s}
The compressibility $\kappa$ and superfluid density $\rho_s$ at finite $L$ and $\beta$ are usually defined in Monte Carlo simulations using the worm algorithm as the variance of the winding numbers $N_x$ and $N_\tau$ such that \cite{Wallin_1994_2DSI}
\begin{align}\label{eq:kappa_def_worm}
    \kappa=\frac{\beta}{L} \left(\langle N_x^2 \rangle -\langle N_x \rangle^2\right),\\
    \rho_s=\frac{L}{\beta} \left( \langle N_\tau^2 \rangle -\langle N_\tau \rangle^2 \right).
\end{align}
However, these definitions yield results which do not converge well in the limit $L,\beta \to \infty$. In Fig.~\ref{fig:kappa_2defs} left, we plot $\kappa$ against the chemical potential $\mu$ for a LL ($g=0$) with $K=0.5$ and $u=1$. It is known that in the $\beta,L \to \infty$ limit, one should have $\kappa = \frac{K}{\pi u}$ and $\rho_s = \frac{K u}{\pi}$, but the previous definitions converge to those results only \emph{after averaging over a small window}, and the oscillations increase as $K$ is lowered. This is why we prefer the more common definitions in the context of bosonization \cite{Giamarchi}
\begin{align}\label{eq:kappa_def_bosonization}
    \kappa &= \frac{q^2}{\pi^2}\langle |\varphi(q,0)|^2 \rangle |_{q=\frac{2\pi}{L}}, \\
    \rho_s &= \frac{\omega_n^2}{\pi^2}\langle |\varphi(0,\omega_n)|^2 \rangle |_{\omega_n =\frac{2\pi}{\beta}},
\end{align}
which display much better convergence properties as shown by $\kappa$ plotted in Fig.~\ref{fig:kappa_2defs} right. The link between both definitions of, for instance $\kappa$, comes from the fact that the Fourier transform of $\partial_x \phi(x,\tau)=\pi N_x/L + \partial_x \varphi(x,\tau)$ at $q$ and $\omega_n=0$ is $\pi N_x \sqrt{\beta/L}\delta_{q,0} + q \varphi(q,0)$. Equation~\eqref{eq:kappa_def_worm} therefore quantifies the fluctuation of the $q=0$ mode, while Eq.~\eqref{eq:kappa_def_bosonization} quantifies that of the smallest non-zero mode $q_{\rm min}=\frac{2\pi}{L}$. The subtle issue at play is thus that both do not coincide in the limit $\beta,L \to \infty$ despite the fact that $q_{\rm min} \to 0$.

\begin{figure}[h!]
    \centering
    \includegraphics[width=0.9\linewidth]{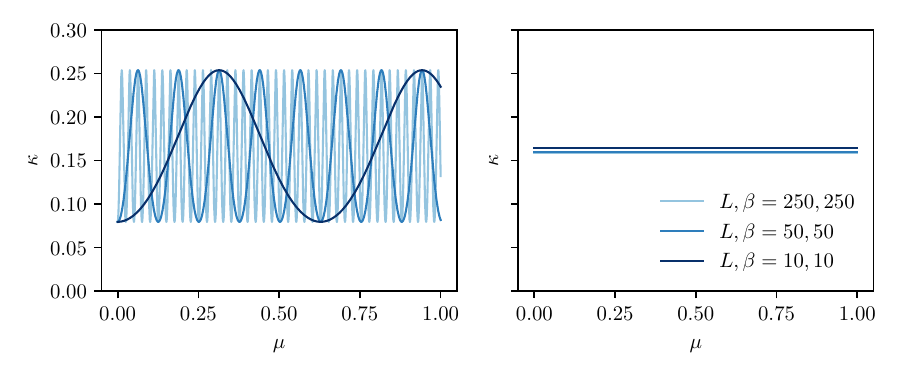}
    \caption{Estimations of the compressibility $\kappa$ obtained for finite $L,\beta$ using Eq.~\eqref{eq:kappa_def_worm} on the left, and Eq.~\eqref{eq:kappa_def_bosonization} on the right, for a LL at $K=0.5$, $u=1$ and varying $\mu$. These plots follow from analytical formulas which are easily obtained since, for $g=0$, the action is quadratic in all degrees of freedom.}
    \label{fig:kappa_2defs}
\end{figure}

\section{Continuous-time Monte Carlo formalism}
\label{app:code}
\subsection{Infinitesimal generator}
A continuous-time Markov process over a configuration space $\Omega=\{x\}$ can be seen as the infinitesimal limit $\Delta t \to 0^+$ of a discrete-time Markov Chain over $\Omega$ with time steps $\Delta t$. The transition matrix ${\cal P}^{\Delta t}$ of such a Markov chain encodes the probability to transition from $x$ to $x'$ through the coefficient ${\cal P}^{\Delta t}(x, x')$. If the time step $\Delta t$ is small enough, we expect the transition matrix to admit the expansion ${\cal P}^{\Delta t}=\mathbbm{1} + \Delta t\mathcal{A}$ where the operator $\mathcal{A}$ is called the (infinitesimal) generator \cite{Davis_1993,Guyon_2023,Monemvassitis_2023}. More formally, the generator is defined as acting on test functions (i.e. observables) as
\begin{align}\label{eq:gen_def}
    \mathcal{A}f(x)=\lim_{\Delta t \to 0^+}\mathbbm{E}_{\varphi_{\Delta t}}\left[\frac{f(\varphi_{\Delta t}(x))-f(x)}{\Delta t}\right],
\end{align}
where $\varphi_{\Delta t}$ is the time-evolution operator over a time $\Delta t$.

Given a target distribution $\pi(x)$, the ergodic theorem states that a discrete-time Markov-Chain will, after some mixing time, sample $\pi(x)$ if it is irreducible, aperiodic and satisfies the stationarity condition (or equivalently the global balance condition)
\begin{align}
    \sum_{x'}  \pi(x') {\cal P}^{\Delta t}(x',x)=\pi(x).
\end{align}
Taking the infinitesimal time step limit $\Delta t \to 0^+$ and introducing a test function, one deduces the global balance condition for the generator
\begin{align}\label{eq:global_balance}
    \sum_x \pi(x) \mathcal{A} f(x)=0.
\end{align}

\subsection{The generator of ECMC algorithms: a simple example}
We now detail how to derive the infinitesimal generator of an ECMC algorithm from a simple example. We consider a particle whose coordinate is $y \in \mathbbm{R}$ with probability $\pi(y) \propto e^{-S(y)}$. We lift the configuration space by adding a speed $e=\pm 1$, and decide that the deterministic motion is
\begin{align}
    \partial_t y(t)=e.
\end{align}
It is interrupted by events that occur with the rate $[e\partial_y S(y)]_+$ and flip $e \to -e$. If the configuration is $(y,e)$ at time $t$, then, for $\Delta t \ll 1$, the configuration at time $t+ \Delta t$ can be
\begin{itemize}
    \item $(y + e \Delta t,e)$ with probability $1- [e\partial_y S(y)]_+ \Delta t$,
    \item $(y,-e)$ with probability $[e\partial_y S(y)]_+ \Delta t$.
\end{itemize}
From definition \eqref{eq:gen_def}, this leads to the generator
\begin{align}
    \mathcal{A}f(y,e)=&\lim_{\Delta t \to 0^+}\frac{\big(1- [e\partial_y S(y)]_+ \Delta t\big)\big(f(y+ e \Delta t,e)-f(y,e)\big) +  [e\partial_y S(y)]_+ \Delta t \big(f(y,-e)-f(y,e)\big)}{\Delta t}\nonumber\\
    =&e \partial_y f(y,e) +  [e\partial_y S(y)]_+  (f(y,-e)-f(y,e)).
\end{align}
It turns out that this generator satisfies the global balance condition \eqref{eq:global_balance} for the stationary distribution $\pi_{\rm tot.}(y,e)=\frac{1}{2}\pi(y)$. Indeed, a small computation shows that
\begin{align}
    \sum_{e=\pm 1} \int \d y\, \pi_{\rm tot.}(y,e) \mathcal{A} f(y,e)=&\sum_{e=\pm 1} \int \d y\, \frac{1}{2}\pi(y) \Big(e \partial_y f(y,e) +  [e\partial_y S(y)]_+  (f(y,-e)-f(y,e))\Big)\nonumber\\
    =&\sum_{e=\pm 1} \int \d y\, \frac{1}{2}\pi(y) \Big(e \partial_y f(y,e) +  ([(-e)\partial_y S(y)]_+ - [e\partial_y S(y)]_+) f(y,e)\Big)\nonumber\\
    =&\sum_{e=\pm 1} \int \d y\, \frac{1}{2}\pi(y) \Big(e \partial_y f(y,e) -e\partial_y S(y) f(y,e)\Big)\nonumber\\
    =&\sum_{e=\pm 1} \int \d y\, \frac{e}{2}\partial_y  \Big(\pi(y) f(y,e)\Big)=0,\nonumber\\.
\end{align}
since $\pi(y)$ vanishes at $\pm \infty$ (otherwise $\pi(y)$ is not normalizable).

\subsection{Proving the global balance condition}
In this section we prove that the SmoWo algorithm introduced in Sec.~\ref{sec:MC_algorithm} is irreducible and satisfied the global balance condition \eqref{eq:global_balance} using the generator notation introduced in the previous subsection. At the end of the section we also discuss the Wo algorithm. Following the discussion of Sec.~\ref{sec:MC_algorithm}, the model is defined over the space of all possible worm heads $p_h$ and tails $p_t$, currents $\vec{J}$, fluctuations $f$, first lifting variables $v=(i,e)$, second lifting variables $\nu=(\alpha,\varepsilon)$, and move types $\sigma$. The SmoWo algorithm samples the variable $(p_h,p_t,\vec{J},f,v,\nu,\sigma)$. Because we have included refreshment updates, it is clear that the algorithm can reach any $v$ and $\nu$ and can update any component of $f$. Moreover, the worm updates generate the entire set of allowed variables $(p_h,p_t,\vec{J})$. Finally, the $\sigma$ updates can update $\sigma$ from any configuration. Putting everything together shows that the algorithm is irreducible. We now prove that the global balance condition is satisfied for the following stationary distribution
\begin{equation}\label{eq:target_distribution}
    \pi_{\rm tot.}(p_h,p_t,\vec{J},f,v,\nu,\sigma)=\pi(p_h,p_t,\vec{J},f) \mu_V(v)\mu_\mathcal{V}(\nu | p_h) \mu_\Sigma(\sigma),
\end{equation}
where $\pi(p_h,p_t,\vec{J},f)=\mathbbm{1}_{\mathcal{C}^2(p_h,p_t)}(\vec{J}) e^{-S(\vec{J},f)}/Z_{\rm w}$ encodes the physical part of the target distribution and all other variables are uniformly distributed over their possible values $\mu_\Sigma(\sigma)=1/2$, $\mu_V(v)=\frac{1}{2L\beta}$ and $\mu_\mathcal{V}(\nu=(\alpha,\varepsilon)|p_h)=\frac{1}{8}\mathbbm{1}_{\partial p_h}(\alpha)$. Following the process outlined in the previous subsection, the SmoWo's generator is found to be (we use the notation $\phi=(p_h,p_t,\vec{J},f)$)
\begin{align}\label{eq:SmoWo_gen}
    \mathcal{A}f(\phi,v,\nu,\sigma)=&(1-\sigma)\Bigg[e\partial_{f_i} f(\phi,v,\nu,0)\nonumber\\
    &\qquad +\sum_{j\in \partial_\q i}[e \partial_{f_i}S_\q^{i,j}(\phi) ]_+ \left(f(\phi,(j,e),\nu,0)-f(\phi,(i,e),\nu,0)\right)\nonumber\\
    &\qquad +[e \partial_{f_i}S_\c^i(\phi) ]_+ \left(f(\phi,(i,-e),\nu,0)-f(\phi,(i,e),\nu,0)\right)\nonumber\\
    &\qquad +\lambda_{\rm r}  \sum_{v'} \mu_V(v') \Big(f(\phi,v',\nu,0) - f(\phi,v,\nu,0) \Big)\nonumber\\
    &\qquad +\lambda_{\rm w} \int \d \phi' \sum_{\nu'} \mathcal{P}(\phi,\nu \to \phi',\nu')\Big( f(\phi',v,\nu',1) - f(\phi,v,\nu,0) \Big)\Bigg]\nonumber\\
    &+\sigma \Bigg[ \varepsilon \partial_{f_\alpha} f(\phi,v,\nu,1)\nonumber\\
    &\qquad +\sum_{\gamma \in \partial_\q \alpha \cap \partial p_h}[\varepsilon \partial_{f_\alpha} S_\q^{\alpha,\gamma}(\phi) ]_+ \left(f(\phi,v,(\gamma,\varepsilon),1)-f(\phi,v,(\alpha,\varepsilon),1)\right)\nonumber\\
    &\qquad +\sum_{\gamma \in \partial_\q \alpha \setminus \partial p_h}[\varepsilon \partial_{f_\alpha} S_\q^{\alpha,\gamma}(\phi) ]_+ \left(f(\phi,v,(\alpha,-\varepsilon),1)-f(\phi,v,(\alpha,\varepsilon),1)\right)\nonumber\\
    &\qquad +[\varepsilon \partial_{f_\alpha}S_\c^\alpha(\phi) ]_+ \left(f(\phi,v,(\alpha,-\varepsilon),1)-f(\phi,v,(\alpha,\varepsilon),1)\right)\nonumber\\
    &\qquad +\lambda_{\rm w} \big( f(\phi,v,\nu,0) - f(\phi,v,\nu,1) \big)\Bigg],
\end{align}
where $\partial_\q i=\{ i+\hat{x}, i-\hat{x}, i+\hat{\tau}, i-\hat{\tau}\}$ is the set of nearest neighbours of $i$, $\mathcal{P}(\phi,\nu \to \phi',\nu')$ is the transition matrix encoding the worm move $\phi \to \phi'$ and the refreshment update $\nu \to \nu'$ if $\phi \ne \phi'$. It can be more explicitly written down in terms of the transition matrix ${\cal P}_{\rm w}(\phi \to \phi')$ for the worm moves alone as
\begin{align}
    \mathcal{P}(\phi,\nu \to \phi',\nu')=\begin{cases}
        \mathcal{P}_{\rm w}(\phi \to \phi)\delta_{\nu,\nu'} &\text{ if } \phi=\phi',\\
        \mathcal{P}_{\rm w}(\phi \to \phi')\mu_\mathcal{V}(\nu'|p_h') &\text{ if } \phi\ne \phi'.
    \end{cases}
\end{align}
Since ${\cal P}_{\rm w}$ was designed to obey the detailed balance condition with respect to the measure $\pi(\phi)$, one can show that ${\cal P}$ satisfies the larger detailed balance condition
\begin{align}\label{eq:P_detailed_balance}
    \pi(\phi')\mu_\mathcal{V}(\nu'|p_h ')\mathcal{P}(\phi',\nu'\to \phi,\nu)=\pi(\phi)\mu_\mathcal{V}(\nu|p_h)\mathcal{P}(\phi,\nu\to \phi',\nu').
\end{align}
To ensure that the total probability flow is conserved, the transition matrix also satisfies the normalization condition
\begin{align}\label{eq:P_normalization}
    \sum_{\nu'}\int \d \phi' \, \mathcal{P}(\phi,\nu\to \phi',\nu')=1.
\end{align}
Equation \eqref{eq:SmoWo_gen} can be split into the first five lines which give the dynamics when $\sigma = 0$, and the last five lines which concern $\sigma=1$. For each of these blocks, the first line is the deterministic shift, the next three are the events which do not change $\sigma$ (including the refreshment term for $\sigma=0$) and the last line is the worm event which tries to move the worm if $\sigma=0$ and always switches $\sigma \to 1-\sigma$. We now prove the global balance condition
\begin{align}\label{eq:global_balance_SmoWo}
    \int \d \phi \sum_{v,\nu,\sigma} \pi_{\rm tot.}(\phi,v,\nu,\sigma) \mathcal{A}f(\phi,v,\nu,\sigma)=0,
\end{align}
following the lines of \cite{Monemvassitis_2023,bouvdup_2025_bosonized1d}. Injecting Eqs.~\eqref{eq:target_distribution} and \eqref{eq:SmoWo_gen} into the global balance condition \eqref{eq:global_balance_SmoWo}, we thus have to show that the following expression vanishes
\begin{align}
    ({\rm i})&(1-\sigma)\Bigg[\int \d \phi \sum_{v,\nu} \pi(\phi) \mu_\mathcal{V}(\nu | p_h) e\partial_{f_i} f(\phi,v,\nu,0)\nonumber\\
    ({\rm ii})&\qquad +\int \d \phi \sum_{v,\nu} \pi(\phi) \mu_\mathcal{V}(\nu | p_h) \sum_{j\in \partial_\q i}[e \partial_{f_i}S_\q^{i,j}(\phi) ]_+ \left(f(\phi,(j,e),\nu,0)-f(\phi,(i,e),\nu,0)\right)\nonumber\\
    ({\rm iii})&\qquad +\int \d \phi \sum_{v,\nu} \pi(\phi) \mu_\mathcal{V}(\nu | p_h) [e \partial_{f_i}S_\c^i(\phi) ]_+ \left(f(\phi,(i,-e),\nu,0)-f(\phi,(i,e),\nu,0)\right)\nonumber\\
    ({\rm iv})&\qquad +\int \d \phi \sum_{v,\nu} \pi(\phi) \mu_\mathcal{V}(\nu | p_h) \lambda_{\rm r}  \sum_{v'} \mu_V(v') \Big(f(\phi,v',\nu,0) - f(\phi,v,\nu,0) \Big)\nonumber\\
    ({\rm v})&\qquad +\int \d \phi \sum_{v,\nu} \pi(\phi) \mu_\mathcal{V}(\nu | p_h) \lambda_{\rm w} \int \d \phi' \sum_{\nu'} \mathcal{P}(\phi,\nu \to \phi',\nu')\Big(f(\phi',v,\nu',1) - f(\phi,v,\nu,0) \Big)\Bigg]\nonumber\\
    ({\rm vi})&+\sigma\Bigg[\int \d \phi \sum_{v,\nu} \pi(\phi) \mu_\mathcal{V}(\nu | p_h) \varepsilon \partial_{f_\alpha} f(\phi,v,\nu,1)\nonumber\\
    ({\rm vii})&\qquad +\int \d \phi \sum_{v,\nu} \pi(\phi) \mu_\mathcal{V}(\nu | p_h) \sum_{\gamma \in \partial_\q \alpha \cap \partial p_h}[\varepsilon \partial_{f_\alpha} S_\q^{\alpha,\gamma}(\phi) ]_+ \left(f(\phi,v,(\gamma,\varepsilon),1)-f(\phi,v,(\alpha,\varepsilon),1)\right)\nonumber\\
    ({\rm viii})&\qquad +\int \d \phi \sum_{v,\nu} \pi(\phi) \mu_\mathcal{V}(\nu | p_h) \sum_{\gamma \in \partial_\q \alpha \setminus \partial p_h}[\varepsilon \partial_{f_\alpha} S_\q^{\alpha,\gamma}(\phi) ]_+ \left(f(\phi,v,(\alpha,-\varepsilon),1)-f(\phi,v,(\alpha,\varepsilon),1)\right)\nonumber\\
    ({\rm ix})&\qquad +\int \d \phi \sum_{v,\nu} \pi(\phi) \mu_\mathcal{V}(\nu | p_h) [\varepsilon \partial_{f_\alpha}S_\c^\alpha(\phi) ]_+ \left(f(\phi,v,(\alpha,-\varepsilon),1)-f(\phi,v,(\alpha,\varepsilon),1)\right)\nonumber\\
    ({\rm x})&\qquad +\int \d \phi \sum_{v,\nu} \pi(\phi) \mu_\mathcal{V}(\nu | p_h) \lambda_{\rm w} \big( f(\phi,v,\nu,0) - f(\phi,v,\nu,1) \big)\Bigg].
\end{align}
As we will see, it turns out that $({\rm i})+({\rm ii})+({\rm iii})=({\rm iv})=({\rm v})+({\rm x})=({\rm vi})+({\rm vii})+({\rm viii})+({\rm ix})=0$. Let us begin by showing that $({\rm i})+({\rm ii})+({\rm iii})=0$. Using the pair-wise symmetry $\partial_{f_i} S^{i,j}_\q(\phi)=-\partial_{f_j} S^{i,j}_\q(\phi)$, the term $({\rm ii})$ becomes
\begin{align}
    ({\rm ii})=&\int \d \phi \sum_\nu \pi(\phi) \mu_\mathcal{V}(\nu | p_h) \sum_{e,i} \sum_{j\in \partial_\q i}[e \partial_{f_i}S_\q^{i,j}(\phi) ]_+ \left(f(\phi,(j,e),\nu,0)-f(\phi,(i,e),\nu,0)\right)\nonumber\\
    =&\int \d \phi \sum_\nu \pi(\phi) \mu_\mathcal{V}(\nu | p_h) \sum_e\left[ \sum_{i,j\in \partial_\q i}[-e \partial_{f_j}S_\q^{i,j}(\phi) ]_+ f(\phi,(j,e),\nu,0)-\sum_{i,j\in \partial_\q i}[e \partial_{f_i}S_\q^{i,j}(\phi) ]_+ f(\phi,(i,e),\nu,0)\right]\nonumber\\
    =&\int \d \phi \sum_\nu \pi(\phi) \mu_\mathcal{V}(\nu | p_h) \sum_e \sum_{i,j\in \partial_\q i}\left( [-e \partial_{f_i}S_\q^{i,j}(\phi) ]_+ - [e \partial_{f_i}S_\q^{i,j}(\phi) ]_+\right) f(\phi,(i,e),\nu,0)\nonumber\\
    =&-\int \d \phi \sum_{v,\nu} \pi(\phi) \mu_\mathcal{V}(\nu | p_h)  \sum_{j\in \partial_\q i} e \partial_{f_i}S_\q^{i,j}(\phi) f(\phi,(i,e),\nu,0)\nonumber\\
    =&-\int \d \phi \sum_{v,\nu} \pi(\phi) \mu_\mathcal{V}(\nu | p_h)  e \partial_{f_i}S_\q(\phi) f(\phi,v,\nu,0),
\end{align}
where $S_\q(\phi)=\sum_{\langle i,j \rangle} S_\q^{i,j}(\phi)$ is the total quadratic action. The term $({\rm iii})$ is computed as 
\begin{align}
    ({\rm iii})=&\int \d \phi \sum_\nu \pi(\phi) \mu_\mathcal{V}(\nu | p_h)\sum_{e,i} [e \partial_{f_i}S_\c^i(\phi) ]_+ \left(f(\phi,(i,-e),\nu,0)-f(\phi,(i,e),\nu,0)\right)\nonumber\\
    =&\int \d \phi \sum_\nu \pi(\phi) \mu_\mathcal{V}(\nu | p_h)\sum_{e,i} \left( [-e \partial_{f_i}S_\c^i(\phi) ]_+ - [e \partial_{f_i}S_\c^i(\phi) ]_+\right)f(\phi,(i,e),\nu,0)\nonumber\\
    =& - \int \d \phi \sum_{v,\nu} \pi(\phi) \mu_\mathcal{V}(\nu | p_h) e \partial_{f_i}S_\c^i(\phi)f(\phi,(i,e),\nu,0)\nonumber\\
    =& - \int \d \phi \sum_{v,\nu} \pi(\phi) \mu_\mathcal{V}(\nu | p_h) e \partial_{f_i}S_\c(\phi)f(\phi,v,\nu,0),
\end{align}
where $S_\c(\phi)=\sum_i S_\c^i(\phi)$ is the total action of the cosine terms. Putting $({\rm i})+({\rm ii})+({\rm iii})$ together and using $\partial_{f_i}\pi(\phi) = -\pi(\phi) \partial_{f_i}\big(S_\q(\phi)+S_\c(\phi)\big)$ thus yields
\begin{align}
    ({\rm i})+({\rm ii})+({\rm iii})=&\int \d \phi \sum_{v,\nu} \pi(\phi) \mu_\mathcal{V}(\nu | p_h) e \left[ \partial_{f_i} f(\phi,v,\nu,0) - \partial_{f_i}S_\q(\phi) f(\phi,v,\nu,0).
     - \partial_{f_i}S_\c(\phi)f(\phi,v,\nu,0)\right]\nonumber\\
     =&\int \d \phi \sum_{v,\nu}  \mu_\mathcal{V}(\nu | p_h) e  \partial_{f_i} \left[ \pi(\phi) f(\phi,v,\nu,0) \right]=0,
\end{align}
since $\pi(\phi)$ vanishes when $f_i \to \pm \infty$ (remember that we relaxed the constraint $f_i\in [1/2,1/2]$ in Sec.~\ref{sec:Wo_algorithm}). Similar computations show that 
\begin{align}
    ({\rm vii})=&-\int \d \phi \sum_{v,\nu} \pi(\phi) \mu_\mathcal{V}(\nu | p_h)   \sum_{\gamma \in \partial_\q \alpha \cap \partial p_h} \varepsilon \partial_{f_\gamma}S_\q^{\alpha,\gamma}(\phi) f(\phi,v,\nu,1)\nonumber\\
    ({\rm viii})=&-\int \d \phi \sum_{v,\nu} \pi(\phi) \mu_\mathcal{V}(\nu | p_h)   \sum_{\gamma \in \partial_\q \alpha \setminus \partial p_h} \varepsilon \partial_{f_\gamma}S_\q^{\alpha,\gamma}(\phi) f(\phi,v,\nu,1)\nonumber\\
    ({\rm ix})=& - \int \d \phi \sum_{v,\nu} \pi(\phi) \mu_\mathcal{V}(\nu | p_h) \varepsilon \partial_{f_\alpha}S_\c(\phi)f(\phi,v,\nu,1).
\end{align}
from which one deduces $({\rm vi})+({\rm vii})+({\rm viii})+({\rm ix})=0$. Since $\mu_V(v)$ is uniform, the refreshment term $({\rm iv})$ is trivially found to vanish
\begin{align}
    ({\rm iv})=&\int \d \phi \sum_{v,\nu} \pi(\phi) \mu_\mathcal{V}(\nu | p_h) \lambda_{\rm r} \sum_{v'} \mu_V(v')\Big(  f(\phi,v',\nu,0) - f(\phi,v,\nu,0) \Big)\nonumber\\
    =&\int \d \phi \sum_\nu \pi(\phi) \mu_\mathcal{V}(\nu | p_h) \lambda_{\rm r} \Big( \sum_{v'} f(\phi,v',\nu,0) - \sum_v f(\phi,v,\nu,0) \Big)=0,
\end{align}
and the only remaining terms are $({\rm v})+({\rm x})$. The term $({\rm v})$ is first rewritten using the detailed balance and normalization conditions (\ref{eq:P_detailed_balance},\ref{eq:P_normalization}) on the transition matrix $\mathcal{P}$.
\begin{align}
    ({\rm v})=& \lambda_{\rm w} \Big(\int \d \phi\, \d \phi' \sum_{v,\nu,\nu'} \pi(\phi) \mu_\mathcal{V}(\nu | p_h) \mathcal{P}(\phi,\nu \to \phi',\nu')f(\phi',v,\nu',1) - \int \d \phi \sum_{v,\nu} \pi(\phi) \mu_\mathcal{V}(\nu | p_h)f(\phi,v,\nu,0) \Big)\nonumber\\
    =& \lambda_{\rm w} \Big(\int \d \phi' \sum_{v,\nu'} \pi(\phi') \mu_\mathcal{V}(\nu' | p_h') f(\phi',v,\nu',1) - \int \d \phi \sum_{v,\nu} \pi(\phi) \mu_\mathcal{V}(\nu | p_h)f(\phi,v,\nu,0) \Big)\nonumber\\
    =& \lambda_{\rm w} \int \d \phi \sum_{v,\nu} \pi(\phi) \mu_\mathcal{V}(\nu | p_h)\Big( f(\phi',v,\nu',1) - f(\phi,v,\nu,0) \Big).
\end{align}
This is nothing but $-({\rm x})$ so $({\rm v})+({\rm x})=0$, which finishes the proof of the global balance condition for the SmoWo algorithm.

The detailed balance condition for the Wo algorithm can be checked using very similar arguments, starting from its generator
\begin{align}\label{eq:Wo_gen}
    \mathcal{A}f(\phi,v)=&e\partial_{f_i} f(\phi,v)\nonumber\\
    &+\sum_{j\in \partial_\q i}[e \partial_{f_i}S_\q^{i,j}(\phi) ]_+ \left(f(\phi,(j,e))-f(\phi,(i,e))\right)\nonumber\\
    &+[e \partial_{f_i}S_\c^i(\phi) ]_+ \left(f(\phi,(i,-e))-f(\phi,(i,e))\right)\nonumber\\
    &+\lambda_{\rm r} \sum_{v'} \mu_V(v') \Big( f(\phi,v') - f(\phi,v) \Big)\nonumber\\
    &+\lambda_{\rm w} \int \d \phi' \mathcal{P}_{\rm w}(\phi \to \phi')\Big(f(\phi',v) - f(\phi,v) \Big),
\end{align}
where there is, of course, no more the lifting variables $\nu$ and $\sigma$, and ${\cal P}_{\rm w}$ is the transition matrix of the worm moves alone.

\section{ECMC in practice}
\label{app:ECMC_in_practice}
\subsection{Computing the event times}
\label{app:event_times}
This section computes the various event times associated to the quadratic interactions, the cosine interaction, the refreshment term, and the worm events. For a generic event triggered by a rate $\lambda_k(t)$, we compute the event time $t_k$ using inversion sampling. One draws $r \sim {\rm ran}([0,1])$ (the uniform distribution over $[0,1]$) and solves
\begin{align}\label{eq:inv_sampl}
    r&=\exp \left[ -\int_0^{t_k} \lambda_k(t) \d t\right].
\end{align}
Note that we need to draw a new variable $r$ for each event rate. We now solve this equation for each event type.

\paragraph{Worm and refreshment events.} The easiest case to solve is that of the refreshment and worm events. Since their rates $\lambda_{\rm r/w}$ are constant, one directly finds
\begin{align}\label{eq:t_rw}
    t_{\rm r/w} = -\frac{\ln r}{\lambda_{\rm r/w}}.
\end{align}

\paragraph{Quadratic events.} Next, considering for instance the quadratic event between $i$ and $i+\hat{x}$, the associated event rate, for $v=(e,i)$, is $[e\partial_{f_i(t)} S^{i,i+\hat{x}}_\q(t)]_+=[e\frac{\pi}{4K}(J^\tau_{p(i)}+f_i(t) - f_{i+\hat{x}})]_+$ where $f_i(t)=f_i+et$ captures the time evolution since the last event. Eq.~\eqref{eq:inv_sampl} therefore becomes
\begin{align}
    r&=\exp \left[ -\int_0^{t_\q^{i,i+\hat{x}}} \left[e\frac{\pi}{4K}(J^\tau_{p(i)}+f_i + et - f_{i+\hat{x}})\right]_+ \d t\right].
\end{align}
Defining $y_\q=e\left(f_i-f_{i+\hat{x}}+J_{p(i)}^\tau\right)$, the previous equation can be rewritten as
\begin{align}
    -\frac{4K}{\pi}\ln r&=\int_0^{t_\q^{i,i+\hat{x}}} [t + y_\q ]_+ \d t=\int_{y_\q}^{t_\q^{i,i+\hat{x}}+y_\q} [t]_+ \d t.
\end{align}
The lower bound of the integral can be replaced by $[y_\q]_+$ since it takes on non-zero values when $t>0$ and $t>y_\q$. The upper bound must be positive since $\frac{\pi}{4K} [t_\q^{i,i+\hat{x}} + y_\q]_+$ corresponds to the event rate which triggered the event. This allows to replace the integrand $[t]_+$ by $t$ and leads to
\begin{align}
    -\frac{4K}{\pi}\ln r&=\frac12 (t_\q^{i,j}+y_\q)^2 - \frac12 [y_\q]_+^2.
\end{align}
which, keeping the solution $t^{i,i+\hat{x}}_\q>0$, gives
\begin{equation}\label{eq:t_q_ij}
    t_\q^{i,i+\hat{x}}=-y_\q + \sqrt{[y_\q]_+^2- \frac{8 K}{\pi}\ln r}.
\end{equation}

\paragraph{Cosine event} The cosine event time $t_\c^i$ occurs with the rate $[e\partial_{f_i(t)}S_\c^i(t)]_+=[e2\pi g\sin(2\pi f_i(t))]_+$ with $f_i(t)=f_i+et$. It is sampled by solving
\begin{align}
    r&=\exp \left[ -\int_0^{t_\c^i} [e2 \pi g \sin(2\pi(f_i + e t))]_+ \d t\right]=\exp \left[ -\int_{e f_i}^{e f_i(t_\c)} 2\pi g[\sin(2\pi x)]_+ \d x\right].
\end{align}
To compute the integral, we decompose the initial and final field as $f_i=e(n + f_{\rm init.})$ and $f_i(t_\c^i) = e(m + f_{\rm fin.})$ with $n,m \in \mathbb{Z}$ and $f_{\rm init.},f_{\rm fin.}\in[0,1[$. This leads to
\begin{align}\label{eq:cos_inv_sampl_inter}
    -\frac{\ln r}{2\pi g}&=\int_{n + f_{\rm init.}}^{m + f_{\rm fin.}} [\sin(2\pi x)]_+ \d x\nonumber\\
    &=\int_{n + f_{\rm init.}}^n [\sin(2\pi x)]_+ \d x+\int_n^m [\sin(2\pi x)]_+ \d x+\int_m^{m + f_{\rm fin.}} [\sin(2\pi x)]_+ \d x\nonumber\\
    &=-\int_0^{f_{\rm init.}} [\sin(2\pi x)]_+ \d x+\frac{m-n}{\pi} +\int_0^{f_{\rm fin.}} [\sin(2\pi x)]_+ \d x.
\end{align}
Since an event occurs at $t_\c^i$, we must have $[e 2\pi g \sin(e 2\pi f_{\rm fin.})]_+ \ne 0$, i.e. $f_{\rm fin.}\in[0,1/2[$, so
\begin{align}\label{eq:cos_inv_sampl1}
    \int_0^{f_{\rm fin.}} [\sin(2\pi x)]_+ \d x=\int_0^{f_{\rm fin.}} \sin(2\pi x) \d x= \frac{1}{2\pi}\left[ 1-\cos (2\pi f_{\rm fin.}) \right].
\end{align}
For $f_{\rm init.}$, there is no similar condition so we write $f_{\rm init.}=(s_\c+ z_\c)/2$ with $s_\c=\lfloor 2 e f_i \rfloor \pmod 2$ the half-integer part, and $z_\c=\{2 e f_i\}$ the rest. Treating $s_c$ as a boolean variable, one writes
\begin{align}\label{eq:cos_inv_sampl2}
    \int_0^{f_{\rm init.}} [\sin(2\pi x)]_+ \d x = (1-s_\c)\int_0^{z_\c/2} \sin(2\pi x) \d x + s_\c \int_0^{1/2} \sin(2\pi x) \d x = \frac{(1-s_\c)}{2\pi}\left[1- \cos(\pi z_\c) \right] + \frac{s_\c}{\pi}.
\end{align}
Plugging Eqs.~(\ref{eq:cos_inv_sampl1},\ref{eq:cos_inv_sampl2}) into Eq.~\eqref{eq:cos_inv_sampl_inter}, one arrives at
\begin{align}
    y_\c &= m-n +\frac{1-\cos (2\pi f_{\rm fin.})}{2}.
\end{align}
where we have defined 
\begin{equation}
    y_\c= -\frac{\ln r}{2 g} + s_\c + (1-s_\c)\frac{1- \cos(\pi z_\c) }{2},
\end{equation}
which only depends on quantities at $t=0$ and is thus known. Taking the integer part and the fractional part of the previous expression yields
\begin{align}
    \lfloor y_\c \rfloor&=m-n,\\
    \{ y_\c \} &= \frac{1-\cos (2\pi f_{\rm fin.})}{2} \Rightarrow f_{\rm fin.} = \frac{1}{2\pi}\arccos(1-2\{ y_\c \}) .
\end{align}
The event time $t_\c^i$ is finally given by
\begin{align}\label{eq:t_c_i}
    t_\c^i = e(f_i(t_\c)-f_i)=m-n + f_{\rm fin.}-f_{\rm init.}=\lfloor y_\c \rfloor +\frac{1}{2\pi}\arccos(1-2\{ y_\c \}) - \{e f_i \}.
\end{align}

\subsection{Pseudocode implementations}
\label{app:pseudo_code}
This section provides detailed pseudocode implementations of the Wo algorithm (Alg.~\ref{alg:Wo}) and SmoWo algorithm (Alg.~\ref{alg:SmoWo}) presented in Sec.~\ref{sec:MC_algorithm}. They both terminate when $n_{\rm sample}$ samples have been outputted. The separate code snippet, Alg.~\ref{alg:sample_output}, computes observables and performs the ballistic motion \eqref{eq:ecmc_motion} until the next event occurs. For $\phi$-dependent observables, we give the example of outputting the entire field $\phi$ at fixed time intervals $T_{\rm sample}$ (think of stroboscopic measurements of a continuous dynamics), but it is, of course, more memory-efficient to only output the observables of interest. As shown in Sec.~\ref{sec:scalar_observables}, we must only output $\phi$-dependent observables when there is no worm, i.e. $p_h=p_t$. For the particle-particle two-point function $C_\theta(p)$, we output the time $t_\theta(p)$ that the algorithm has spent while having $p_h-p_t=p$. The two-point function is then retrieved through $C_\theta(p)=t_\theta(p)/t_\theta(0)$, as argued in Sec.~\ref{sec:theta_2point_function}.
\end{widetext}

\begin{algorithm}
\caption{Ballistic motion $+$ Sampling}\label{alg:sample_output}
{\bf Input} $T_{\rm Sample},t_{\rm event},t_s,i,e,f,\vec{J},p_h,p_t,{\rm Sample},t_\theta$\hspace{-0.11cm} \;
\uIf(\tcp*[f]{Output before event}){$t_s<t_{\rm event}$}{
    $f_i \gets f_i + e \,t_s$\;
    \If{$p_h = p_t$}{
        Rebuild $\phi$ from $f,\vec{J}$ using Eqs.~(\ref{eq:n_f_def},\ref{eq:current_def})\;
        ${\rm Sample}\gets {\rm Sample}\cup \{\phi\}$\;
        $n_{\rm sample} \gets n_{\rm sample}-1$\;
        }
        $f_i \gets f_i + e \,(t_{\rm event}-t_s)$\;
        $t_s\gets t_s - t_{\rm event} + T_{\rm Sample}$ \tcp*{time till next sample output}
    }
    \Else{$t_s\gets t_s- t_{\rm event}$\;
        $f_i \gets f_i + e \,t_{\rm event}$\;}
    $t_\theta(p_h-p_t)\gets t_\theta(p_h-p_t) + t_{\rm event}$\;
{\bf Return} ${\rm Sample}$\;
\end{algorithm}

\begin{algorithm}
\caption{Wo algorithm}\label{alg:Wo}
{\bf Input} $f, \vec{J}, i, e, p_h, p_t, n_{\rm sample}, T_{\rm Sample}, \lambda_{\rm r}, \lambda_{\rm w}$ \;
${\rm Sample}=\{\}$\;
$t_\theta(p)=0$ for all $p$\;
$t_s \gets T_{\rm Sample}$\tcp*{Time till sampling}
\While{$n_{\rm sample}>0$}{
    $j_\q \gets \underset{j\in \partial_\q i}{\rm argmin}(t_\q^{i,j} \gets$ Eq.~\eqref{eq:t_q_ij})\;
    $t_\q \gets \underset{j\in \partial_\q i}{\rm min}(t_\q^{i,j} \gets$ Eq.~\eqref{eq:t_q_ij})\;
    $t_\c \gets$ Eq.~\eqref{eq:t_c_i}\;
    $t_{\rm r} \gets$ Eq.~\eqref{eq:t_rw}\;  
    $t_{\rm event}={\rm min}(t_\q,t_\c,t_{\rm r},t_{\rm w})$\;
    ${\rm Sample} \gets$ Alg.~\ref{alg:sample_output}($\cdots,i,e,\cdots$) \;
    \uIf(\tcp*[f]{Quadratic event}){$t_\q =t_{\rm event}$}{
        $e,i \gets e,j_\q$\;}
    \uElseIf(\tcp*[f]{Cosine event}){$t_\c =t_{\rm event}$}{      
        $e,i \gets -e,i$ \;}
    \uElseIf(\tcp*[f]{Refreshment}){$t_{\rm r} =t_{\rm event}$}{ 
        $i\gets {\rm choice}(\llbracket 1, L\rrbracket\times \llbracket 1, \beta \rrbracket)$\;
        $e\gets {\rm choice}(\{-1,1\})$\;}
    \ElseIf(\tcp*[f]{Worm event}){$t_{\rm w} =t_{\rm event}$}{
        \uIf{$p_h=p_t$ and ${\rm ran}(0,1)<1/2$}{
            $p_h \gets {\rm Choice}(\llbracket 1, L\rrbracket\times \llbracket 1, \beta \rrbracket)$ \;
            $p_t \gets p_h$\;}
        \Else{ $p_h' \gets {\rm Choice}(\{p_h \pm \hat{x},p_h \pm \hat{\tau}\})$\;
            \If{ ${\rm ran}(0,1)<P(p_h \to p_h')$}{
            $p_h \gets p_h'$\;}}}
}
{\bf Return} ${\rm Sample}$\;
\end{algorithm}

\vfill

\begin{algorithm}
\caption{SmoWo algorithm}\label{alg:SmoWo}
{\bf Input} $f, \vec{J}, i, e, \alpha, \varepsilon, \sigma, p_h, p_t, n_{\rm sample}, T_{\rm Sample}$, $\lambda_{\rm r}, \lambda_{\rm w}$ \;
${\rm Sample}=\{\}$\;
$t_\theta(p)=0$ for all $p$\;
$t_s \gets T_{\rm Sample}$\;
\While{$n_{\rm sample}>0$}{
    $t_{\rm w} \gets$ Eq.~\eqref{eq:t_rw}\; 
    \uIf{$\sigma= 0$}{
        $j_\q \gets \underset{j\in \partial_\q i}{\rm argmin}(t_\q^{i,j} \gets$ Eq.~\eqref{eq:t_q_ij})\;
        $t_\q \gets \underset{j\in \partial_\q i}{\rm min}(t_\q^{i,j} \gets$ Eq.~\eqref{eq:t_q_ij})\;
        $t_\c \gets$ Eq.~\eqref{eq:t_c_i}\;
        $t_{\rm r/ w} \gets$ Eq.~\eqref{eq:t_rw}\;
        $t_{\rm event}={\rm min}(t_\q,t_\c,t_{\rm r},t_{\rm w})$\;
        ${\rm Sample} \gets$ Alg.~\ref{alg:sample_output}($\cdots,i,e,\cdots$)\;
        \uIf(\tcp*[f]{Quadratic event}){$t_\q =t_{\rm event}$}{
            $e,i \gets e,j_\q$\;}
        \uElseIf(\tcp*[f]{Cosine event}){$t_\c =t_{\rm event}$}{ 
            $e,i \gets -e,i$\;}
        \uElseIf(\tcp*[f]{Refreshment}){$t_{\rm r} =t_{\rm event}$}{ 
            $i,e \gets {\rm choice}(\llbracket 1, L\rrbracket\times \llbracket 1, \beta \rrbracket \times \{-1,1\})$\;}
        \ElseIf(\tcp*[f]{Worm event}){$t_{\rm w} =t_{\rm event}$}{
            \uIf{$p_h=p_t$ and ${\rm ran}(0,1)<1/2$}{
                $p_h \gets {\rm Choice}(\llbracket 1, L\rrbracket\times \llbracket 1, \beta \rrbracket)$ \;
                $p_t \gets p_h$\;
                $\alpha , \varepsilon \gets {\rm choice}(\partial p_h \times \{-1,1\})$\;}
            \Else{ $p_h' \gets {\rm Choice}(\{p_h \pm \hat{x},p_h \pm \hat{\tau}\})$\;
                \If{ ${\rm ran}(0,1)<P(p_h \to p_h')$}{
                $p_h \gets p_h'$\;
                $\alpha, \varepsilon  \gets {\rm choice}(\partial p_h \times \{-1,1\})$\;}}
            $\sigma \gets 1$\;}
        }
    \Else{
        $\gamma_\q \gets \underset{\gamma \in \partial_\q \alpha}{\rm argmin}(t_\q^{\alpha,\gamma} \gets$ \hspace{-0.2cm} Eq.~\eqref{eq:t_q_ij} with $v \to \nu$)\;
        $t_\q \gets \underset{\gamma \in \partial_\q \alpha}{\rm min}(t_\q^{\alpha,\gamma} \gets$ Eq.~\eqref{eq:t_q_ij} with $v \to \nu$)\;
        $t_\c \gets$ Eq.~\eqref{eq:t_c_i} with $v \to \nu$\;
        $t_{\rm event}={\rm min}(t_\q,t_\c,t_{\rm w})$\;
        ${\rm Sample} \gets$ Alg.~\ref{alg:sample_output}($\cdots,\alpha,\varepsilon,\cdots$)\;
        \uIf(\tcp*[f]{Quad.}){$t_\q =t_{\rm event}$ and $\gamma_\q \in \partial p_h$}{
            $\varepsilon,\alpha \gets \varepsilon, \gamma_\q$\;}
        \uElseIf{$t_\q =t_{\rm event}$ and $\gamma_\q \notin \partial p_h$}{ 
            $\varepsilon,\alpha \gets -\varepsilon, \alpha$\;}
        \uElseIf(\tcp*[f]{Cosine event}){$t_\c =t_{\rm event}$}{ 
            $\varepsilon,\alpha \gets -\varepsilon, \alpha$\;}
    \ElseIf(\tcp*[f]{Worm event}){$t_{\rm w} =t_{\rm event}$}{
        $\sigma \gets 0$\;}
    }}
{\bf Return} ${\rm Sample}$\;
\end{algorithm}

\clearpage

\bibliography{refs}

\end{document}